\documentclass[twocolumn]{aastex62}
\usepackage{amsmath} 

\graphicspath{{./}{figures/}}

\received{2020 March 9}
\revised{2020 April 20}
\accepted{2020 April 20}
\submitjournal{ApJ}

\newcommand{\simname}[1]{\texttt{#1}}

\begin{document}
\title{Outbursts in Global Protoplanetary Disk Simulations}

\correspondingauthor{Kundan Kadam}
\email{kundan.kadam@csfk.mta.hu}

\author[0000-0002-0786-7307]{Kundan Kadam}
\affiliation{Konkoly Observatory, Research Center for Astronomy and Earth Sciences, Konkoly-Thege Mikl\'{o}s \'{u}t 15-17, 1121 Budapest, Hungary}

\author{Eduard Vorobyov}
\affiliation{Department of Astrophysics, The University of Vienna, A-1180 Vienna, Austria}
\affiliation{Ural Federal University, 51 Lenin Str., 620051 Ekaterinburg, Russia}
\author{Zsolt Reg\'{a}ly}
\affiliation{Konkoly Observatory, Research Center for Astronomy and Earth Sciences, Konkoly-Thege Mikl\'{o}s \'{u}t 15-17, 1121 Budapest, Hungary}
\author{\'{A}gnes K\'{o}sp\'{a}l}
\affiliation{Konkoly Observatory, Research Center for Astronomy and Earth Sciences, Konkoly-Thege Mikl\'{o}s \'{u}t 15-17, 1121 Budapest, Hungary}
\affiliation{Max Planck Institute for Astronomy, K\"onigstuhl 17, D-69117 Heidelberg, Germany}
\affiliation{ ELTE E\"otv\"os Lor\'and University, Institute of Physics, P\'azm\'any P\'eter s\'et\'any 1/A, 1117 Budapest, Hungary}

\author{{P}\'{e}ter \'{A}brah\'{a}m}
\affiliation{Konkoly Observatory, Research Center for Astronomy and Earth Sciences, Konkoly-Thege Mikl\'{o}s \'{u}t 15-17, 1121 Budapest, Hungary}
\affiliation{ ELTE E\"otv\"os Lor\'and University, Institute of Physics, P\'azm\'any P\'eter s\'et\'any 1/A, 1117 Budapest, Hungary}


\begin{abstract}
While accreting through a circumstellar disk, young stellar objects are observed to undergo sudden and powerful accretion events known as FUor or EXor outbursts. 
Although such episodic accretion is considered to be an integral part of the star formation process, the triggers and mechanisms behind them remain uncertain. 
We conducted global numerical hydrodynamics simulations of protoplanetary disk formation and evolution in the thin-disk limit, assuming both magnetically layered and fully magnetorotational instability (MRI)-active disk structure.
In this paper, we characterize the nature of the outbursts occurring in these simulations. 
The instability in the dead zone of a typical layered disk results in ``MRI outbursts". 
We explore their progression and their dependence on the layered disk parameters as well as cloud core mass. 
The simulations of fully MRI-active disks showed an instability analogous to the classical thermal instability. 
This instability manifested at two temperatures--above approximately 1400 K and 3500 K--due to the steep dependence of Rosseland opacity on the temperature. 
The origin of these thermally unstable regions is related to the bump in opacity resulting from molecular absorption by water vapor and may be viewed as a novel mechanism behind some of the shorter duration accretion events.
Although we demonstrated local thermal instability in the disk, more investigations are needed to confirm that a large-scale global instability will ensue. 
We conclude that the magnetic structure of a disk, its composition, as well as the stellar mass, can significantly affect the nature of episodic accretion in young stellar objects.
\end{abstract}


\section{Introduction} 
\label{sec:intro}

During the process of star formation, the angular momentum contained in the parent molecular cloud results in an accretion disk surrounding the star, which ultimately provides material for planet formation.
Several lines of reasoning suggest that the accretion through such a protostellar (early times, relatively massive disk) or protoplanetary disk is not steady and monotonically decreasing but highly time dependent.
Embedded protostars in star forming regions consistently show about an order of magnitude lower luminosities as compared to simple theoretical predictions \citep{KH1995}.
An assumption that the young star mostly accretes in a low state, which is punctuated by periods of high accretion rates, can solve this luminosity problem.
Such variations, if they occur during late stages of YSO evolution, may also explain the luminosity spread observed in the Hertzsprung--Russell diagram of young clusters \citep{Hillenbrand2009}. 
Observations of young stellar objects (YSOs) indeed show powerful and long-lived outbursts; these are called the FU Orionis (FUor) or EX Lupi (EXor) class of objects \citep[][]{HK1996,Herbig2008, Audard2014}.
FUors show outbursts with an amplitude of several magnitudes, corresponding to the luminosity of about $100 L_{\odot}$, and lasting for up to over a hundred years.
EXors typically exhibit about an order of magnitude weaker and short-duration outbursts, although recent observations suggest that there may be a continuum of duration and amplitudes rather than two distinct classes.
{Thus, episodic accretion is considered to be an integral part of low-mass star formation, occurring at all early stages and responsible for about $1/10$ of the final stellar mass \citep{DV12}.}
A number of models have been put forward to explain the origin of this eruptive behavior, including thermal instability \citep{BL1994}, gravo-magneto instability \citep{Armitage2001}, binary interactions \citep{BB1992}, accretion of protoplanetary clumps \citep{VB2005}, and magnetic field--disk interactions \citep{DS2010}, including stellar magnetic cycles \citep{Armitage2016}. 
{Despite of this progress, as we elaborate in the following sections,} there remain significant uncertainties concerning the mechanisms behind the outbursts.
{ The implications of time-dependent accretion in YSOs also need to be taken into account in all theories of planet formation, as these events can have profound effects on disk dynamics, chemistry, mineralogy, dust properties, and snow lines.}


In this paper, we concentrate on the mechanisms behind luminosity outbursts that were modeled in our simulations presented in \cite{Kadam2019}.
We conducted global hydrodynamic simulations of protoplanetary disk formation and evolution in the thin-disk limit, where the infall from the contracting cloud core and a sub-astronomical-unit inner computational boundary were included.
The simulations are performed with two distinct assumptions of the vertical disk structure. 
A magnetically layered disk is assumed in the first set of simulations, where magnetorotational instability (MRI) is supported only in the surface layers of the disk due to external sources of ionization. 
The second set of simulations is conducted with the assumption that the protoplanetary disk is MRI active throughout.
In \cite{Kadam2019}, we showed that a magnetically layered disk forms a dead zone, and moreover, near the inner edge of the dead zone, axisymmetric dynamic rings also form due to the action of viscous torques.
A ring migrates inward as the large-scale spiral density waves associated with gravitationally unstable regions carry away angular momentum.
Such a ring ultimately undergoes an instability, resulting in the rapid accretion of the accumulated material onto the star and a corresponding luminosity outburst.
Because these outbursts are associated with the sudden activation of magnetorotational instability, they are termed as ``MRI outbursts".
{Here the layered disk was modeled with an effective \cite{SS1973} $\alpha$ parameter and considered to be fully active above a specified critical ionization temperature, while nonideal magnetohydrodynamic effects were ignored.}
The fully MRI active disks, which are still used for observational modeling, show a fundamentally different type of outburst related to the thermal instability of the disk.
Classical thermal instability ensues at high temperatures when hydrogen is ionized, due to the rapid change of opacity with temperature. 
In such conditions, the cooling is decreasing with temperature and the disk fails to maintain a vertical thermal equilibrium.  
We show that similar instability can occur at much lower temperatures when sophisticated gas opacities, especially the effects of molecular absorption by water vapor, are taken into account.
{Although the simulations do not take into account dust growth and dynamics, the effects of dust are considered implicitly via opacities with an adopted dust-to-gas ratio of $1\%$.}
We elaborate on the nature of eruptions related to both MRI outbursts and thermal instability in detail, focusing on a typical instance of each case.
The simulations also show accretion outbursts related to gravitational fragmentation and migration of the clumps onto the protostar; we do not concentrate on this mechanism as it has been extensively studied, e.g., by \cite{VB2015}  and \cite{VE2018}.

In Section \ref{sec:model}, we briefly summarize the disk model and the simulations considered in this paper.
In Section \ref{sec:mri}, we investigate the trigger and progression of an MRI outburst occurring in magnetically layered disks, while in Section \ref{sec:TI}, we investigate the nature of thermal instability in fully MRI active disks.
We summarize and conclude our results in Section \ref{sec:conclusions}.

\section{Model Description and Initial Parameters}
\label{sec:model}

We solve the full set of numerical hydrodynamic equations for the evolution of the circumstellar disk in the thin-disk limit with a cylindrical geometry.
This approach has the advantage of realistically modeling non-axisymmetric structures in the disk while being computationally feasible for a global and long-term evolution compared to full three-dimensional simulations. 
The equations for the conservation of mass, momentum, and energy are solved, while the pressure is calculated via the ideal equation of state.
The cooling rate of the disk is based on the analytical solution of the radiation transfer equations in the vertical direction \citep{Dong2016}.
{The heating includes background irradiation, as well as stellar photospheric and accretion luminosity irradiation, where the flaring of the disk is also taken into account \citep{VB2010}. }
The Planck and Rosseland mean opacities needed to compute cooling and heating are taken from \citet{Semenov2003}.
The coevolution of the forming protostar is calculated using the stellar evolution tracks provided by the \simname{STELLAR} code \citep{Yorke2008}.
We use a carefully implemented inflow--outflow (transparent) boundary condition at the inner boundary, where the gas is allowed to flow freely from the computational domain to the sink cell and vice versa \citep{Vorobyov2018}.
{The inner boundary was placed at 0.42 au so that the behavior of the innermost, sub-astronomical-unit-scale scale regions of the disk can be captured.}
The numerical simulations start from the gravitational collapse of a starless cloud core and evolve into the embedded phase of star formation when a star and the circumstellar disk are formed.
We assume the initial surface density and angular velocity for an axially symmetric core collapse of the nascent cloud core \citep{Basu1997}.
For a more detailed description of the model including equations and initial conditions, see \cite{Kadam2019}.

We now summarize our model of accretion through a disk with vertical magnetic structure. 
The physical reasons behind the layered accretion and formation of a dead zone in the disk, as well as the mechanism behind MRI outbursts, are elaborated in Section \ref{sec:mri}.
The kinematic viscosity is parametrized using the \cite{SS1973} $\alpha$ prescription:
\begin{equation}
     \nu=\alpha_{\rm eff} c_{\rm s} H
\end{equation}
where $c_{\rm s}$ is the sound speed and $H$ is the vertical scale height.
The accretion through a layered disk structure is taken into account by considering an effective and adaptive $\alpha$ parameter: 
\begin{equation}
\alpha_{\rm eff}=\frac{\Sigma_{\rm a} \alpha_{\rm a} + \Sigma_{\rm d} \alpha_{\rm d}}{\Sigma},
\label{eq:alpha}
\end{equation}
where $\Sigma_{\rm a}$ is the gas surface density of the MRI-active layer at the disk surface, $\Sigma_{\rm d}$ is that of the MRI-dead layer at the midplane, and $\Sigma= \Sigma_{\rm a} + \Sigma_{\rm d}$ is the total surface density \citep{Bae2014}.
The parameters $\alpha_{\rm a}$ and $\alpha_{\rm d}$ set the strength of turbulent viscosity in the MRI-active and MRI-dead layers of the disk, respectively. 
We set $\alpha_{\rm a}$ to a canonical value of 0.01, while $\alpha_{\rm d}$ is defined as 
\begin{equation}
\alpha_{\rm d} = \alpha_{\rm MRI,d} + \alpha_{\rm rd},
\end{equation}
where
\begin{equation}
\alpha _{\rm MRI,d} =
\left\{
\begin{array}{c}
\alpha_{\rm a}, \,\, \mathrm{if} \,\, T_{\rm mp} > T_{\rm crit}   \\
0, \,\,\, \mathrm{otherwise},
\label{MRI_dead}
\end{array}
\right. 
\end{equation}
and $T_{\rm mp}$ is the disk midplane temperature, while $T_{\rm crit}$ is the MRI activation temperature due to the thermal effects caused by alkaline metals or dust sublimation.
Thus, with equation (\ref{MRI_dead}), the disk is assumed to switch rapidly from a magnetically dead to a magnetically active state above $T_{\rm crit}$.
Magnetohydrodynamics simulations show that the dead zone can have some nonzero residual viscosity due to hydrodynamic turbulence driven by the Maxwell stress in the active layer \citep{Okuzumi2011, Bai2016}. 
Therefore, we define the residual viscosity as
\begin{equation}
\alpha_{\rm rd} = \min \left( 10^{-4} , \alpha_{\rm a} \frac{\Sigma_{\rm a}}{\Sigma_{\rm d}} \right).
\label{eq:alphaRD}
\end{equation}
This expression ensures that accretion in the dead zone cannot exceed that of the active zone, because the nonzero $\alpha_{\rm rd}$ is assumed to be caused by the turbulence propagating from the
active layer down to the disk midplane. 

Another important aspect of the disk physics in terms of the eruptive behavior is the properties of dust grains and gas species taken into account as sources of disk opacities. 
We have implemented the Rosseland and Planck mean opacities based on the models of \cite{Semenov2003}, spanning temperatures from $5$ K and $10000$ K and gas densities between $10^{-18}{\rm g\,cm^{-1}}$ and $10^{-7}{\rm g\,cm^{-1}}$, which cover the entire range for protoplanetary disks around low-mass stars.
{Dust grains are the primary source of opacity in almost the entire disk, although above the dust sublimation temperature of about $1500$ K, line and continuum absorption as well as scattering 
due to molecular species becomes important.}
The temperatures near the inner boundary of the disk do not significantly exceed this value for a layered disk model. 
However, in a fully magnetically active disk, represented by a constant value of $\alpha_{\rm a} = 0.01$ in the simulations, the inner regions get much hotter in comparison due to viscous heating. 
{We will show that such a disk exhibits self-regulating thermal instability outbursts associated with the steep dependence of opacity on temperature--in particular, the opacity caused by molecular lines of water vapor shapes these regions of instability.}

\begin{table*}
\caption{List of Simulations}
\label{table:sims}
\begin{tabular}{|l|}
\hline
\begin{tabular}{p{2.2cm}p{1.5cm}p{1.3cm}p{1.3cm}p{1.5cm}p{2.5cm}}
\hspace{-1.2cm} Model Name & \hspace{-0.4cm} ${ \rm M_{gas} (M_{\odot}) }$ & \hspace{-0.3cm} ${\rm \beta} (\times 10^{-3})$   & \hspace{-0.2cm} ${\rm T_{crit} (K)}$ & \hspace{-0.4cm} ${\rm \Sigma_{\rm a}}$ (g~cm$^{-2}$) & \hspace{-0.2cm}  { $M_{\rm *,F}(M_\odot)$}\\
  \end{tabular}\\ \hline
\begin{tabular}{l}
$\kern-\nulldelimiterspace\left.
  \begin{tabular}{p{3.2cm}p{1.5cm}p{1.3cm}p{1.5cm}p{1.5cm}p{1.5cm}}
\hspace{-0.7cm}\simname{model1\_T1300\_S100}$^1$   & 1.152 &  1.360  & 1300 & 100 &0.74  \\ 
\hspace{-0.7cm}\simname{model1\_T1300\_S10}    & 1.152 &  1.360  & 1300 & 10 &0.65  \\ 
\hspace{-0.7cm}\simname{model1\_T1500\_S100}   & 1.152 &  1.360  & 1500 & 100  &0.74\\
\hspace{-0.7cm}\simname{model2\_T1300\_S100}   & 0.346 &  1.355  & 1300 & 100 & 0.28 \\
  \end{tabular}\right\}$ Magnetically layered disk
\\ 
\end{tabular}  
\\ \hline
\begin{tabular}{l}
$\kern-\nulldelimiterspace\left.
  \begin{tabular}{p{3.2cm}p{1.5cm}p{1.3cm}p{1.5cm}p{1.5cm}p{1.5cm}}
\hspace{-0.7cm}\simname{model2\_const\_alpha}  & 0.346 &  1.355  & -- & -- & 0.24\\
\hspace{-0.7cm}\simname{model2\_const\_alpha\_0.2au}  & 0.346 &  1.355  & -- & -- & --
  \end{tabular}\right\}$ Fully MRI-active disk
\end{tabular}  
\\
\hline
\end{tabular}\\
    \vspace{1ex}
     \small
      $^1$ Fiducial layered disk model
\end{table*}

Table \ref{table:sims} lists the simulations that are considered in this paper for the analysis of the outbursts. 
We study one representative instance of an MRI outburst that occurred in the fiducial model presented in \cite{Kadam2019}, i.e., in \simname{model1\_T1300\_S100}.
Here, the total mass of the gas in the initial prestellar cloud core {($M_{\rm gas}$)} was set to 1.152 $M_\odot$, while the ratio of the rotational to the gravitational energy {($\beta$)} was consistent with the observations of prestellar cores.
We qualitatively demonstrated the effect of the model parameters $\Sigma_{\rm a}$ and $T_{\rm crit}$ as well as the initial core mass, with the help of three additional simulations -- \simname{ model1\_T1300\_S10}, \simname{model1\_T1500\_S100} and \simname{model2\_T1300\_S100}, respectively.
To demonstrate the occurrence of thermal instability, \simname{model2\_const\_alpha} is chosen, with a lower initial prestellar core mass of 0.346 $M_\odot$. 
{ The table also lists the ``final" stellar mass ($M_{\rm *,F}$) defined at 0.45 Myr.
The inner boundary was placed at 0.42 au, with the exception of \simname{model2\_const\_alpha\_0.2au} where the inner boundary was moved to 0.21 au in order to investigate the nature of thermal instability.
This simulation was only conducted up to 0.1 Myr, and thus its final mass is not listed.} 
Note that we defer a quantitative, statistical analysis of the burst phase to a subsequent paper when all simulations have progressed past the { classical T Tauri phase} and the disks show no more eruptions.
The resolution of all simulations was $512 \times 512$ in the radial and azimuthal directions, with logarithmic and uniform spacing, respectively. 
The highest grid resolution in the radial direction was $8.2\times10^{-3}$ au at the innermost cell of the disk, while the poorest resolution near the outer boundary of the cloud (at $10,210$ au) was about 202 au for the fiducial model.

\section{MRI Outburst}
\label{sec:mri}
In this section we first explain the mechanism behind an MRI outburst and then elaborate on its general properties with the help of an example case from the fiducial layered disk model. We will also show the effects of model parameters on such outbursts.

\subsection{Mechanism}

The mechanism behind MRI outbursts is inevitably linked with the layered accretion occurring in protoplanetary disks.
MRI is considered to be the primary source of viscosity in a protoplanetary disk in its non-self-gravitating regions \citep{BH1991,HGB1995}.
{Because MRI causes turbulence due to the shearing Keplerian motion of ionized gas coupled to a weak magnetic field,
the disk needs to have a sufficient level of ionization for its sustenance.}
Galactic cosmic rays are considered to be the major source of ionization, penetrating a nearly uniform gas surface density
of the disk surface \citep{UN1981}.
In our model, this effect is parameterized by the constant thickness of the active layer, $\Sigma_{\rm a}$.
Additional sources such as stellar far-ultraviolet or X-ray radiation, and radioactive decay of short-lived isotopes are typically much weaker in comparison at a few astronomical units from the star \citep{Turner2014}.
Collisional ionization of alkali metals 
is expected to occur at a certain temperature (corresponding to $T_{\rm crit}$ in our model), which can cause a sudden, almost exponential rise of the electron fraction in the gas.
This condition is typically satisfied only in the innermost 0.1-1 au of the disk, making this region fully MRI active.
{Beyond about 15 au, because of the disk's low gas surface density as well as flaring geometry, the aforementioned external sources can penetrate the entire disk column. 
This provides sufficient ionization for the MRI to operate, making the outer disk fully turbulent.}
A protoplanetary disk thus accretes through a layered disk structure in the inner regions, where the mass transfer primarily occurs in the ionized surface layers, while a magnetically dead zone forms in the midplane \citep{Gammie1996}. 
In the presence of powerful stellar (or disk) winds or a strong magnetic environment, it is possible that the disk is shielded from the galactic cosmic rays \citep{ST1968, Padovani2011}.
Despite these uncertainties and additional complications due to nonideal magnetohydrodynamical effects, most theories predict the formation of a dead zone with low effective viscosity at the distance of a few astronomical units from the star \citep{Armitage2011}.

The dead zone essentially forms a bottleneck in the angular momentum transport, and material can accumulate in its vicinity due to the mismatch in mass transfer rate.
\cite{Gammie1996} proposed that the buildup of the material at the boundary of the dead zone will ultimately result in sufficient viscous heating to raise the temperature above $T_{\rm crit}$.
Then, the disk can abruptly change its state from having a dead zone to being fully magnetically active.
One-dimensional models indeed showed that layered accretion results in quasi-steady behavior of protoplanetary disks, with accretion outbursts separated by quiescent intervals \citep{Armitage2001,Zhu2010a,Martin2012}. 
The mechanism was termed gravo-magneto (GM) instability, as the heating from the viscosity due to self-gravity is responsible for triggering MRI. 
The increased viscosity results in the accretion of the accumulated material onto the star, causing an accretion outburst.
Recent studies show that the dead zone is not entirely dead; several pure hydrodynamic instabilities, such as vertical shear instability, convective overstability, zombie vortex instability, and vertical shear instability, should cause a small, residual viscosity \citep[e.g.,][]{Nelson2013,KH2014,Marcus2015,SK16}. 
With the inclusion of such viscosity in the dead zone, the MRI is triggered at the inner boundary of the disk due to viscous heating in the dead zone, resulting in an inside-out burst, as opposed to an outside-in burst with the GM instability \citep{Bae2013}.

The outburst models mentioned so far are implemented in one-dimensional approximation, and intrinsically non-axisymmetric behavior such as vortices is ignored, while the effects of gravitational instability, large-scale spirals, and clumps are parameterized using a local viscosity.  
\cite{Bae2014} conducted two-dimensional simulations of MRI instability, however, their model had a few limitations.  
Their model for vertical disk structure was similar to that used for our simulations, with a few key differences. The disk was considered MRI active only when the azimuthally averaged midplane temperature exceeded $T_{\rm crit}$.
Thus, MRI was triggered at a given radius throughout the disk simultaneously.
Their initial model started with a disk surrounding a protostar with an ad hoc surface density distribution, with the outer boundary of the disk placed at 100 au.
They also implemented a 10\% azimuthally asymmetric density fluctuation of the infalling material, without which the simulations did not generate the gravitational instability (GI)-induced spiral waves responsible for accretion bursts.

With the above limitations in mind, in \cite{Kadam2019} we presented a set of simulations to investigate the formation and evolution of protoplanetary disks accreting through the magnetically layered structure.
As described in Section \ref{sec:model} the simulations started with the collapse phase of the parent molecular cloud, thus including the disk formation and early T Tauri stages of evolution.
We noted the effects of model parameters ${T_{\rm crit}}$ and ${\rm \Sigma_{\rm a}}$ as well as those of the initial mass of the cloud core, and contrasted the findings with fully magnetically active disks.
We showed that during the evolution of a layered disk of a sun-like star, a dead zone forms in the inner $\approx${15} au region with the canonical values of the parameters (${ T_{\rm crit}}=1300$~K and ${\Sigma_{\rm a}}=100$~g\,cm$^{-2}$, i.e., the fiducial layered disk model -- \simname{model1\_T1300\_S100}).
This dead zone was not uniform, and near its inner boundary at a few astronomical units, long-lived, axisymmetric, high surface density rings are formed due to the action of viscous torques.
These rings were characterized by high surface density and low effective viscosity.
The rings showed complex, dynamical behavior; multiple gaseous rings could form simultaneously and showed merger as well as vortex formation.
A ring migrated inward faster than the local viscous timescale due to the angular momentum carried away by large-scale spiral waves.
These waves were generated because the accumulated gas in the rings was marginally gravitationally unstable.
As compared to the smooth surface density structure of a fully MRI-active disk, the rings formed in the magnetically layered disks showed remarkable contrast in the conditions of the inner disk. 

\subsection{Outburst Analysis}
\label{sub:analysis}

\begin{figure}
\centering
\includegraphics[width=0.5\textwidth]{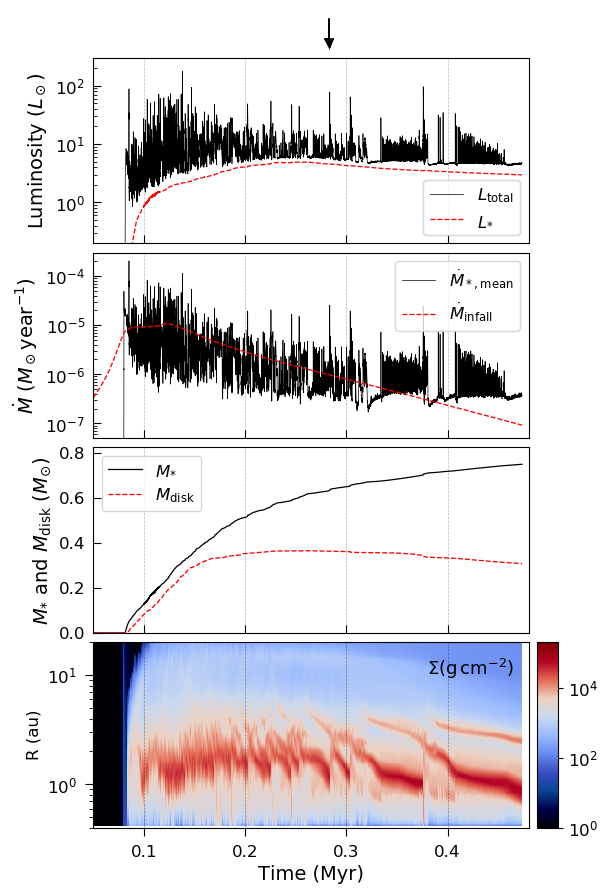}
\caption{Time-dependent quantities -- total and stellar luminosities, mean mass accretion rate, stellar and disk mass -- as well as spacetime diagram of the surface density of the inner 20 au region for the fiducial model, \simname{model1\_T1300\_S100}.
The arrow shows the particular eruption studied in detail.
}
\label{fig:model1alltime}
\end{figure}

Figure \ref{fig:model1alltime} shows the evolution of global quantities as well as that of the disk surface density for \simname{model1\_T1300\_S100}. 
Here, the evolutionary time starts from the onset of gravitational collapse, i.e., at the start of the simulation. 
The first panel shows the stellar ($L_{*}$) and total ($L_{\rm total}$) luminosities, where the former is calculated through the stellar evolution code and the latter is calculated as the sum of stellar and accretion luminosities. 
The protostar is formed at about 0.09 Myr, and thus after this point, these quantities can be measured.
The second panel in Figure \ref{fig:model1alltime} shows the evolution of the mean mass accretion rate onto the central protostar ($\dot{M}_{\rm *, mean}$), as well as the envelope infall rate ($\dot{M}_{\rm infall}$).
The accretion rate onto the protostar is calculated from the amount of gas passing through the inner computational boundary. 
The accretion rate is averaged backwards in time with a rolling window of 10 and 40 yr, above and below an instantaneous value of $\dot{M}_{\rm *}=10^{-5} {\rm M_\odot}$ respectively. 
This averaging smooths out the stochastic variations in the turbulent flow at the inner boundary. 
The mass of the material passing to the sink cell through the sink--disk interface is redistributed between the central protostar and the sink cell 
with a ratio of 
95\%:5\%.
For numerical stability, the inner boundary condition is switched from outflow only (from disk to sink) to inflow--outflow only after the formation of the protostar.
This switch causes an artificial peak in the mass transfer rate, which is noticeable in Figure \ref{fig:model1alltime}, and is ignored in the analysis of all models.  
The mass transport rate at 500 au is used as a proxy for the envelope infall rate for the disk.   
The third panel shows the mass of the protostar ($M_{*}$) and that of the disk ($M_{\rm disk}$).
{ The disk remains a substantial fraction of the mass of the central protostar throughout the evolution, ending up at about 40\% of its mass after 0.5 Myr.}
{At the end of the simulation, the system can be considered to be in a T Tauri phase, when most of the envelope has accreted onto the central star and the disk.}

\begin{figure}
\centering
\includegraphics[width=0.49\textwidth]{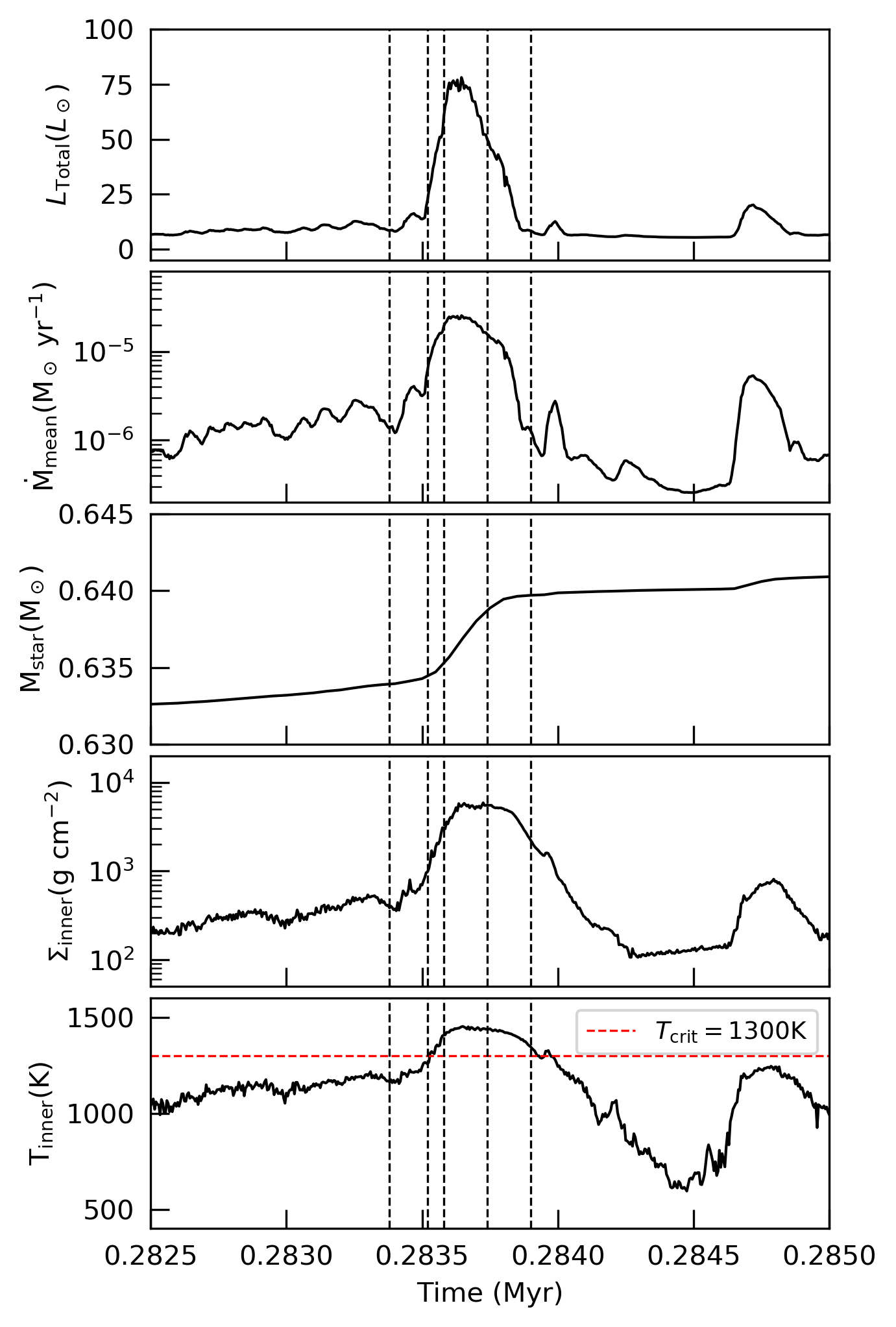}
\caption{Time-dependent system properties (total luminosity, $\dot{M}_{\rm mean}$, stellar mass) as well as surface density and midplane temperature at the inner boundary during the MRI outburst occurring in \simname{model1\_T1300\_S100}. The vertical dashed lines mark the time corresponding to the 2D plots presented in Figures \ref{fig:mri2D} and 1D profiles in Figure \ref{fig:mri1Dpro}, while the red dashed line shows $T_{\rm crit}=1300$~K.}
\label{fig:mriLight}
\end{figure}

After disk formation until about 0.4 Myr, $\dot{M}_{\rm *, mean}$ approximately matches the steadily declining $\dot{M}_{\rm infall}$.
The continuing infall from the envelope as well as the massive disk ensures the inner disk has sufficient supply of mass coming in from the outer region at all times.
Several strong accretion events coinciding with the increase in the total luminosity is noticeable throughout the evolution. 
The strongest of these events increases the mass transfer rate from about $10^{-6}$ to over $10^{-5} {\rm M_{\odot}\,year^{-1}}$ and the luminosity from a few ${\rm L_{\odot}}$ to few tens of ${\rm L_{\odot}}$.
Some of the outbursts result from the migration of a fragment onto the star; such clumps form when the disk is massive and susceptible to local gravitational instability and fragmentation. 
{The small-amplitude fluctuations in the mass accretion rate in this early stage arise from the propagation of spiral density waves due to the complex disk structure.}
The last panel of Figure \ref{fig:model1alltime} depicts the spacetime diagram of the disk surface density for the inner 20 au region.
The high surface density rings can be seen as diagonal patterns.
In \cite{Kadam2019}, we stated that FUor-like accretion outbursts occur at the discontinuities in the rings in the spacetime diagram, which corresponds to the accretion of the accumulated material onto the central star. 
We now investigate a representative MRI outburst occurring at about 0.283 Myr, indicated by an arrow at the top in Figure \ref{fig:model1alltime}. 
Note that the MRI outbursts across time as well as with different parameters of the layered disk model are qualitatively similar.

\begin{figure*}
{\bf \hspace{0.75cm} Pre-outburst \hspace{1cm} $\rightarrow$ \hspace{2.25cm} During outburst  \hspace{2.25cm} $\rightarrow$ \hspace{1cm} Post-outburst}\\
${\color{white}\overbrace{Quit playing games with my heart}}$ $\overbrace{{\color{white}my heart, my heart, I should have known from the start}}$ \\
\begin{tabular}{l}
\vspace{-0.22cm}\includegraphics[width=0.985\textwidth]{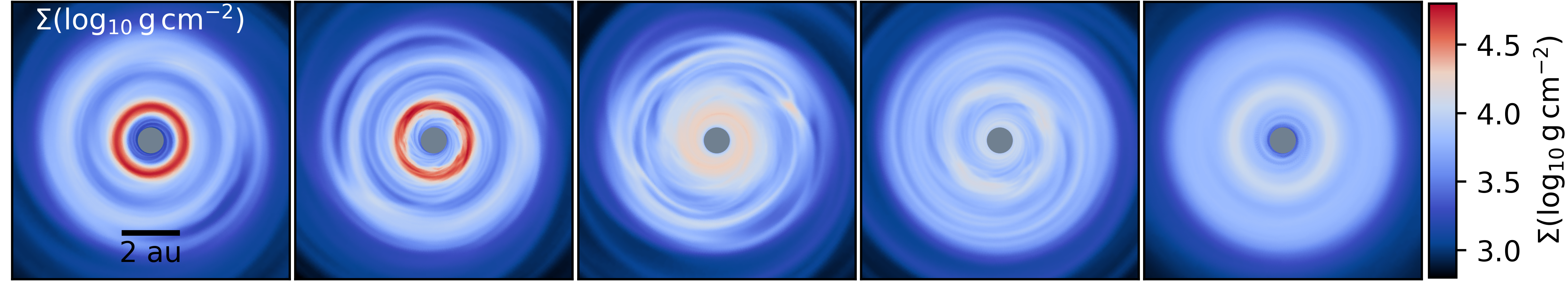} \\ 
\vspace{-0.22cm}\includegraphics[width=\textwidth]{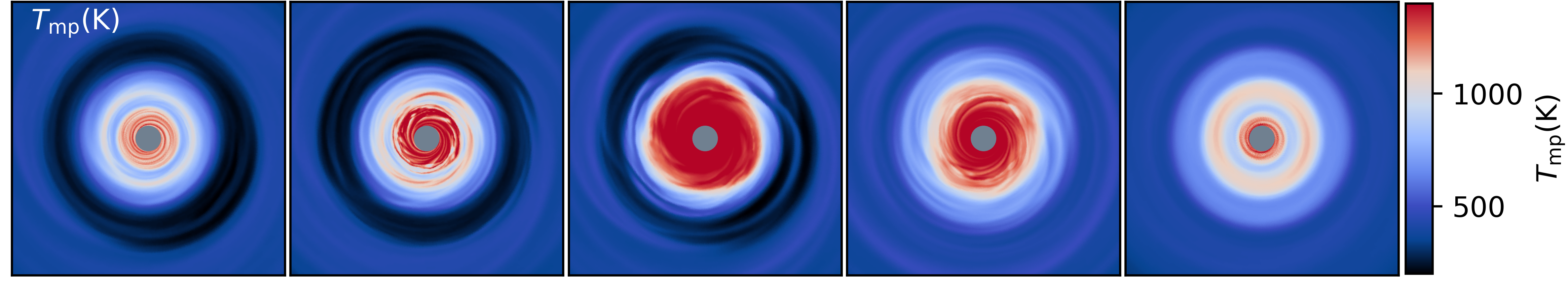} \\
  \includegraphics[width=0.993\textwidth]{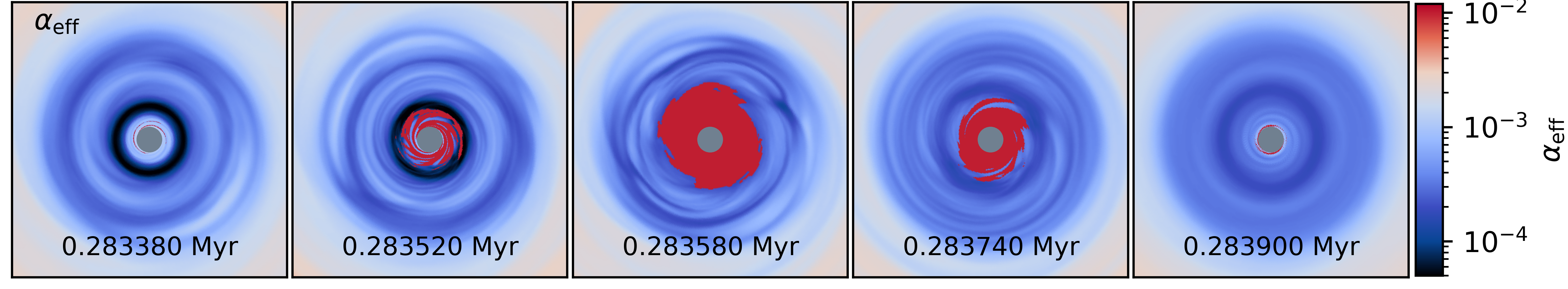} \\
\end{tabular}
\caption{The temporal behavior of the gas surface density, midplane temperature and $\alpha_{\rm eff}$ in the inner $10\times10$ au box of the layered disk, showing the progression of the MRI outburst.} 
\label{fig:mri2D}
\end{figure*}

Due to constraints on data storage, the time resolution of the output of the two-dimensional fields was 2000 yr. 
In order to investigate the outburst progression in detail, we restarted the simulation from a suitable point, and obtained outputs with a finer time resolution of 20 yr.
Figure \ref{fig:mriLight} expands on the 2500 yr period in the vicinity of the MRI outburst indicated in Figure \ref{fig:model1alltime}.
This is a typical MRI outburst and identified on the basis of discontinuity in the innermost ring in the spacetime diagram of the surface density.
The vertical lines in Figure \ref{fig:mriLight} correspond to slices of time that are analyzed in detail in the following analysis.
The first panel in this figure shows the outburst profile in the total luminosity of the system, which increased to about 75 $L_{\odot}$, with a marginally asymmetrical rise and fall. 
The increase occurred over approximately 100 yr while the total duration was about 300 yr, thus, the eruption was slow rising, and qualitatively similar to the observed V1515 Cyg eruption \citep{HK1996}. 
The second panel in Figure \ref{fig:mriLight} shows the evolution of the mass accretion rate onto the central protostar, which increased from about $10^{-6}$ to few times $10^{-5} M_{\odot}{\rm year^{-1}}$ during the outburst.
The third panel shows the mass of the central protostar, which increases by about 1.1\%.  
The changes in $\dot{M}_{\rm*, mean}$ and total luminosity as well as the duration and the total amount of mass accreted during the outburst are consistent with the observed estimates for FUor eruptions \citep{Audard2014}.
The last two panels show the azimuthally averaged surface density and midplane temperature at the inner boundary, i.e., at 0.42 au.
The surface density increased substantially by about an order of magnitude, while the temperature crossed $T_{\rm crit}=1300$ K at the onset of the outburst and remained above this value throughout the duration.

Figure \ref{fig:mri2D} depicts the development of the outburst in terms of gas surface density distribution, midplane temperature, and $\alpha_{\rm eff}$ over the innermost $10\times10$ au region of the disk. 
These snapshots of 2D distributions, or ``frames", are marked in Figure \ref{fig:mriLight} by the vertical dashed lines.
The first frame corresponds to 0.283380 Myr of evolutionary time, and the successive frames are plotted after the first one at the intervals of 140, 200, 360 and 520 yr, respectively.
The frames were chosen such that the disk behavior is captured during the rise, maximum, and decline of the luminosity, as well as before and after the outburst.
The first column in Figure \ref{fig:mri2D} shows the typical ring structure, also elaborated in \cite{Kadam2019}. 
The rings are characterized by high surface density and low viscosity. 
The second column captures the disk structure at the very beginning of the outburst when fractures in the inner ring are noticeable. 
The temperature increased starting at the inner boundary of the disk, and the $\alpha_{\rm eff}$ also increased to the maximum possible value of $\alpha_{\rm a}=0.01$, indicating MRI activation in this region.
The third column at 0.283580 Myr shows the distribution near the peak of both the total luminosity and mass accretion rate.
At this stage, the maximum extent of the progression of the MRI activation, up to approximately 2 au, was observed.
The boundaries of the high-temperature region were smoother than that in $\alpha_{\rm eff}$, as the latter is activated similarly to a step function above the critical temperature.
This MRI-activated region fully engulfed the inner ring and the increased viscosity transported this accumulated material inward, resulting in the large mass transfer rate. 
The fourth column shows the declining phase when the MRI activation front receded inward.
As the mass was depleted from within the inner disk region, the midplane temperature dropped due to decreasing viscous heating and the MRI front followed suit. 
Notice that throughout the outburst, the MRI front showed spiral boundaries and substructures, primarily reflecting the non-axisymmetric activation of the MRI as well as the cooler gas moving in as a result of steady accretion from the outer parts.
In the last frame, which is plotted after the outburst, the inner ring completely disappeared. 
About 0.005 ${M_{\odot}}$ or 5 Jupiter masses of gas, mostly accumulated in the reservoir of the inner ring, was accreted onto the protostar.

\begin{figure}
\centering
\includegraphics[width=0.48\textwidth]{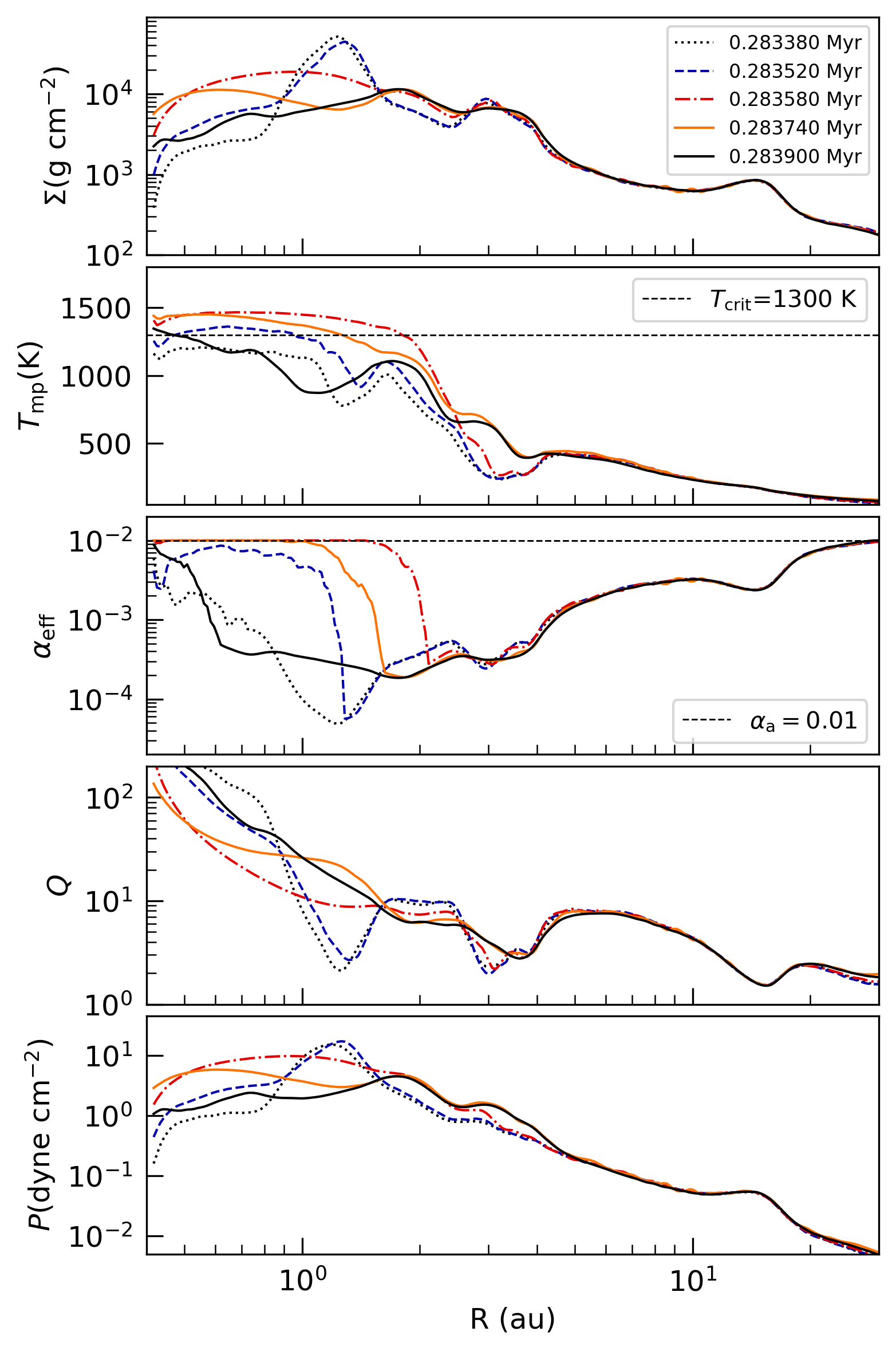}
\caption{Progression of the azimuthally averaged profiles of surface density, midplane temperature, $\alpha_{\rm eff}$, Q-parameter and vertically averaged pressure during the MRI outburst.
}
\label{fig:mri1Dpro}
\end{figure}

\begin{figure}
\centering
  \includegraphics[width=0.46\textwidth]{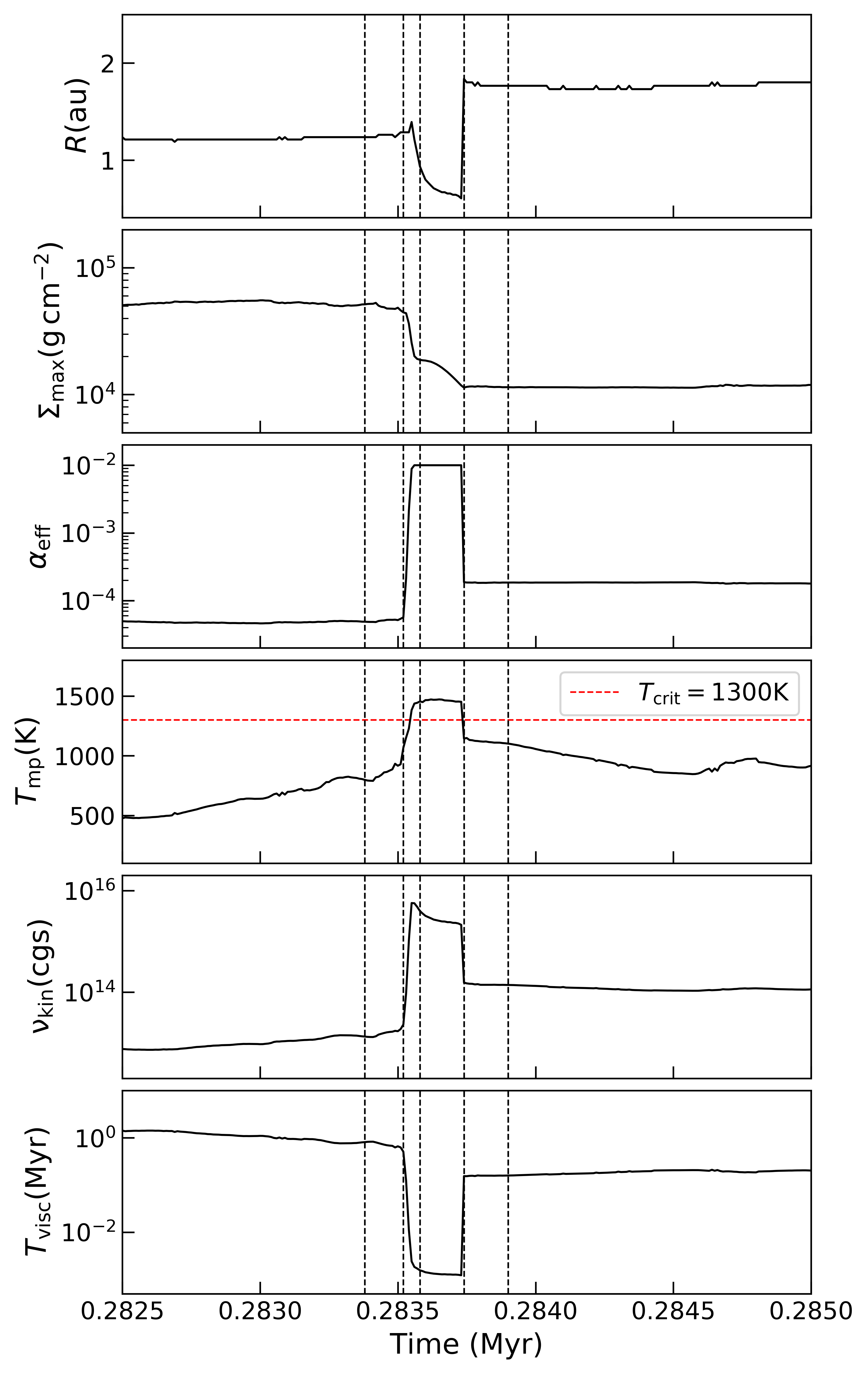}
\caption{
Properties of the innermost ring are plotted as it evolved across the MRI outburst occurring in \simname{model1\_T1300\_S100}.
The first panel shows its location, defined as the radius of the maximum surface density at a given time.
The remaining panels show $\Sigma_{\rm max}$, $\alpha_{\rm eff}$, $T_{\rm mp}$, {kinematic viscosity} and viscous timescale at this location. The vertical dashed lines mark the time corresponding to the 2D plots presented in Figure \ref{fig:mri2D}.}
\label{fig:ringquant}
\end{figure}

Figure \ref{fig:mri1Dpro} shows the azimuthally averaged profiles of the quantities -- surface density, $\alpha_{\rm eff}$, midplane temperature, Toomre $Q$ parameter and vertically averaged pressure -- in the inner 30 au of the disk during the MRI outburst.
Note that the abscissae are in logarithmic units. 
These profiles correspond to the vertical lines in Figure \ref{fig:mriLight} as well as frames presented in Figure \ref{fig:mri2D}.
Two rings can be seen before the outburst at the radius of approximately 1.2 and 3 au, with the inner ring being about an order of magnitude more prominent in terms of the surface density. 
The inward movement of the surface density as well as complete disappearance of the inner ring can be seen as the outburst progressed. 
{A third and much weaker ring can be seen at about 18 au near the outer edge of the dead zone, which remains unperturbed throughout this time.}
The second panel shows the evolution of $T_{\rm mp}$; the temperature front clearly originated near the inner boundary of the disk and moved outward.
As the midplane temperature crossed $T_{\rm crit}$, the ionization front in terms of $\alpha_{\rm eff}$ followed this advancement, reaching its maximum value of $\alpha_{\rm a}=0.01$.
Note that as the activation of MRI above $T_{\rm crit}$ is assumed to be a step function (Equation \ref{MRI_dead}), the azimuthally averaged $\alpha_{\rm eff}$ always lagged behind the $T_{\rm mp}$ profile in its radial progression. 
The origin of the outburst near the inner boundary and its outward advancement are consistent with the ``inside-out" burst models.
With the inclusion of dead zone residual viscosity, even when the magnitude of $\alpha_{\rm rd}$ is small, the MRI is triggered in the innermost region due to the action of viscous heating in the dead zone \citep{Bae2013}. 
{The ionization or MRI front decelerates as it propagates in the outward direction in a ``snowplow" fashion.
This leads to a slow-rising mass accretion rate and a similarly blunt increase in the accretion luminosity \citep{Lin85}.
In contrast, an inward-traveling front will produce an ``avalanche" effect, achieving a sharply rising light curve.}
The $Q$ parameter in the fourth panel of Figure \ref{fig:mri1Dpro}, defined as $Q=c_{\rm s} \Omega/  \pi G \Sigma$ in the usual notation, quantifies the significance of the self-gravity \citep{Toomre64}.
The self-gravity is typically significant at larger radii; however, due to the large amount of gas accumulated in the vicinity of the rings, this region is marginally unstable, showing $Q\approx2$.
The last panel shows the vertically averaged pressure profiles.
Local maxima in pressure are noticeable before the outburst at the location of the rings, which can act as dust traps and accumulate solid material. 
These pressure maxima are strongly perturbed across the outburst.
The azimuthal profiles of all the quantities remained unperturbed in the outer region, beyond about 5 au.

In order to probe the conditions at the inner ring during the outburst, similar to \cite{Kadam2019}, we define a ring to be the structure corresponding to the location of the maximum surface density at a given time.
Note that multiple rings may be simultaneously present at any given time. However, this method should obtain an insight into the local quantities at the inner, most prominent ring, which is relevant across the MRI outburst.
In Figure \ref{fig:ringquant} we consider the evolution of the quantities at the location of the inner ring between 0.2825 and 0.285 Myr.
This time interval is identical to that considered in Figure \ref{fig:mriLight}, with the vertical lines marking the frames corresponding to Figure \ref{fig:mri2D}.
Figure \ref{fig:ringquant} is similar to Figure 7 in \cite{Kadam2019}, where now we have sufficient time resolution to study the behavior of the disk during the outburst. 
The first row shows the radial location of the ring. 
On this timescale, the inward migration of the ring before the outburst, as may be inferred from Figure \ref{fig:model1alltime}, is not noticeable.  
The sudden inward drift that started at about 0.2835 Myr corresponds to the MRI getting triggered at the ring location. 
The discontinuity near the end of the outburst at about 0.2837 Myr occurs because at this point, the inner ring has lost sufficient mass for its surface density to decrease below the maximum value.
The maximum in surface density thus corresponds to the next prominent ring. 
The rate of inward drift of the ring, and by proxy the maximum in the pressure, during the outburst was about 4500 au~Myr$^{-1}$.
This is over two orders of magnitude faster than the inward migration rate of the rings at approximately 25 au~Myr$^{-1}$.

The next three rows of Figure \ref{fig:ringquant} show the maximum in surface density, $\alpha_{\rm eff}$, and midplane temperature at the location of the ring respectively. 
The initial sharp decline in $\Sigma_{\rm max}$ is because of the dissipation of the steep density gradients of the ring after the sudden increase in viscosity.
After the surface density profile was flattened out, the rate of dissipation was proportionally slowed down. 
The qualitative evolution of the sharply peaking ring into a blunt profile can also be inferred from 
Figures \ref{fig:mri2D} and \ref{fig:mri1Dpro}.      
The triggering of MRI can be seen as the sharp rise in $\alpha_{\rm eff}$ to the maximum possible value of $\alpha_{\rm a}=0.01$. 
As the midplane temperature stayed above $T_{\rm crit}$ throughout the development of the outburst, $\alpha_{\rm eff}$ also stayed in high state over this period.
Comparing the midplane temperature at the location of the ring to that at the inner boundary (Figure \ref{fig:mriLight}), the inside-out progression of the MRI front can be inferred. 
The last two rows in Figure \ref{fig:ringquant} depict the evolution of kinematic viscosity and viscous timescale at the ring location. 
The viscosity increased sharply at the onset of the outburst and decreased sharply at the end of it. 
Its magnitude is given by $\nu_{\rm kin}= \alpha_{\rm eff} c_{\rm s}^2/ \Omega_{\rm k} $. 
Because the first two terms are almost constant through the outburst ($c_{\rm s}$ is proportional to $T_{\rm mp}$), the decrease in $\nu_{\rm kin}$ during the outburst reflects the inward drift of the ring.
The viscous timescale, $T_{\rm visc} = r^2/\nu_{\rm kin}$, decreased over three orders of magnitude during the outburst.
The viscous timescale during the outburst is consistent with its total duration of about 300 yr.

\subsection{Effect of Model Parameters}

\begin{figure}
\centering
\includegraphics[width=0.5\textwidth]{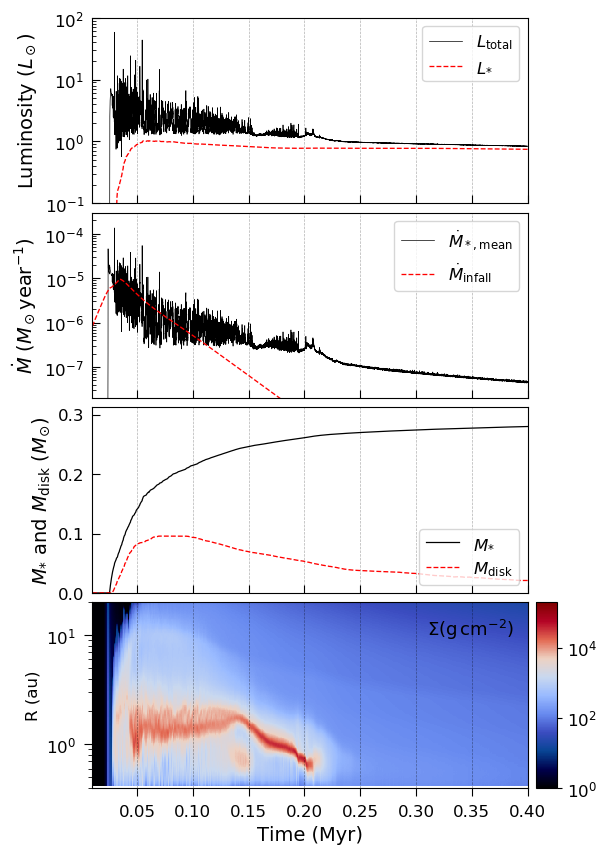}
\caption{Time-dependent quantities -- total and stellar luminosities, mean mass accretion rate, stellar and disk mass -- as well as spacetime diagram of the surface density of the inner 20 au region for the low-mass model, \simname{model2\_T1300\_S100}.}
\label{fig:model3alltime}
\end{figure}

We analyze the effects of the decrease in the stellar mass with the simulation \simname{model2\_T1300\_S100}.
This is the fiducial low-mass model in \cite{Kadam2019}, with the total cloud gas mass of 0.346 $M_\odot$, ultimately producing a fully convective star. 
Figure \ref{fig:model3alltime} shows the evolution of global, time-dependent quantities and the spacetime diagram for this model.
The total luminosity shows fluctuations for the initial duration of about 0.2 Myr, and then it approaches that of the central protostar.
The evolution of mass accretion can be followed past when the infall from the parent cloud has stopped. 
In the early phase, the mass accretion rate showed fluctuations related to the stochastic variations while accreting through a massive disk and associated propagation of spiral waves.
As seen in the spacetime diagram of the surface density in the last panel, the ring formed at about 0.15 Myr and migrated inward.
However, this ring was soon accreted onto the central protostar, and successive rings did not form.
Due to the lower protostellar mass and consequently insufficient viscous heating, the midplane temperature of the disk never reached $T_{\rm crit}$ after the formation of the ring.
The ring eventually viscously dissipated after getting too close to the inner boundary.
The infall had already stopped at this point and the mass in the outer disk reservoir was insufficient to sustain further ring formation. 
Thus, the mass of the cloud core profoundly affected the outbursting behavior of the disk, where, although a high surface density ring was formed, no MRI bursts were observed. 
{The absence of MRI outbursts in this model indicates that there may be a lower limit for the stellar mass, below which FUor eruptions by this mechanism do not occur.}

\begin{figure}
\centering
\includegraphics[width=0.5\textwidth]{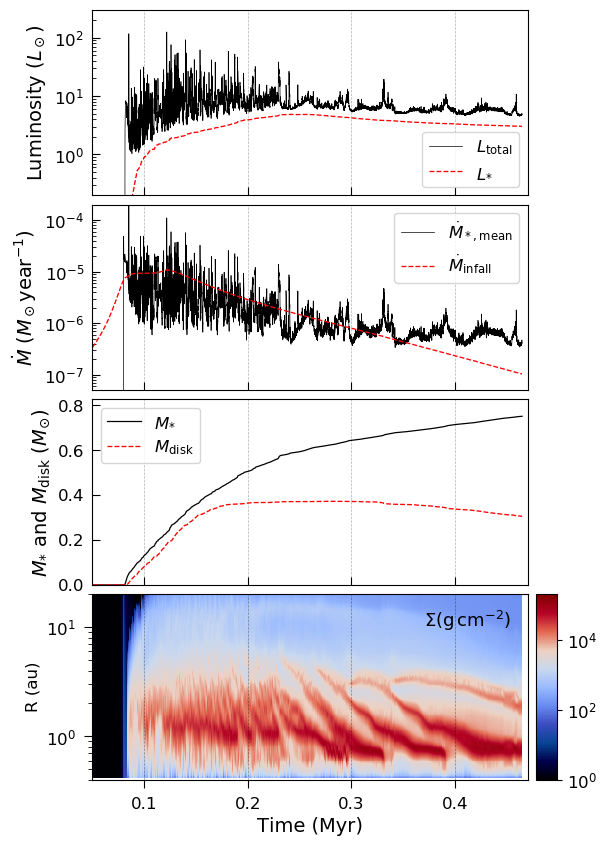}
\caption{Time-dependent quantities -- total and stellar luminosities, mean mass accretion rate, stellar and disk mass -- as well as spacetime diagram of the surface density of the inner 20 au region for \simname{model1\_T1500\_S100}.}
\label{fig:model1Talltime}
\end{figure}

\begin{figure}
\centering
\includegraphics[width=0.5\textwidth]{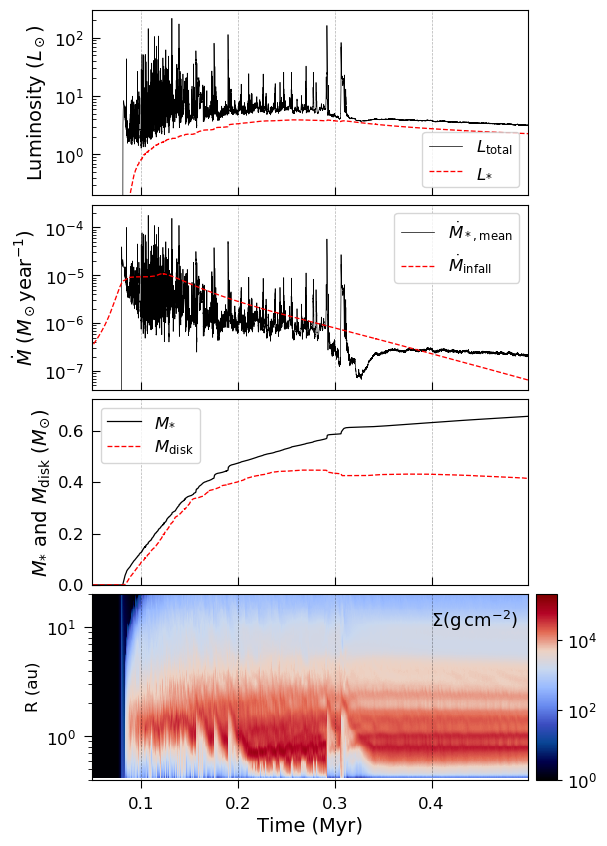}
\caption{Time-dependent quantities -- total and stellar luminosities, mean mass accretion rate, stellar and disk mass -- as well as spacetime diagram of the surface density of the inner 20 au region for \simname{model1\_T1300\_S10}.
}
\label{fig:model1Salltime}
\end{figure}

The effect of MRI activation temperature is studied using \simname{model1\_T1500\_S100}, where $T_{\rm crit}$ is set to $1500$~K.
The spacetime diagram in Figure \ref{fig:model1Talltime} shows the formation of complex ring structures which qualitatively look similar to those formed in the fiducial model with $T_{\rm crit} = 1300$ K.
These rings also migrate inward and seem to eventually disappear; however, there is one key difference.
In the spacetime diagram, the discontinuity when a ring disappeared was not abrupt but spread over time.
The corresponding mass accretion rate as well as the luminosity burst were also much diffused, e.g., as observed at about 0.39 Myr.
Analysis similar to that performed in Section \ref{sub:analysis} revealed 
that during these accretion events, the disk temperature never reached the set critical temperature.
As a consequence, the rings migrated inward and eventually viscously dissipated after getting close to the inner boundary.
Thus, an increase in $T_{\rm crit}$ significantly affected the outbursting behavior of the disks in our simulations.
Here, the rings were formed; however, MRI bursts were not produced and the rings directly crossed the inner boundary.
The exact characterization of such accretion events needs further analysis. 

The uncertainty in the thickness of the active layer is substantial, and $\Sigma_{\rm a}$ can have a much lower value if the disk is shielded from cosmic rays by the magnetic environment of the disk \citep{Cleeves2013}.
Figure \ref{fig:model1Salltime} depicts the effects of decreasing active layer thickness to 10 ${\rm g~cm^{-2}}$. 
As described in \cite{Kadam2019}, in this case a more effective bottleneck is formed in the angular momentum transport.
Thus, the rings formed are more massive and show much slower inward migration.
This initial period was dominated by occasional MRI bursts, as can be seen in the discontinuities in the spacetime diagram,  
while after about 0.35 Myr of evolution, the mass accretion rate stabilized to a few times $10^{-7}$.
The MRI outbursts associated with the discontinuities in the spacetime diagram were more powerful than the fiducial case.
This was because of the larger amount of accumulated mass in the inner disk region.
The signatures of the outburst are also seen in the mass of the star, which increased abruptly during these events. 
The disk remained at a substantial fraction of the stellar mass at the end of the simulation.
Both the continual formation of the rings and the large disk mass suggest that MRI outbursts will continue to occur at later times. 
Gradual radiative cooling may, however, decrease the disk temperature, thus making the MRI bursts less likely.
{ From Table \ref{table:sims}, the final protostellar mass of 0.65 $M_\odot$ was notably smaller than that for the fiducial layered disk model, again indicating a significant impediment to accretion through the disk with a lower $\Sigma_{\rm a}$.}
A more quantitative analysis of the outbursting phase will be performed after the entire ``ring phase" of the disk evolution had been simulated.

\section{Thermal Instability}
\label{sec:TI}

The classical thermal instability was originally proposed to explain the observed outbursts in cataclysmic variables, i.e., dwarf novae in disks around accreting white dwarf systems \citep[e.g.,][]{Patterson84, Lasota2001}.
The same mechanism was adapted for FUors as for outbursts resulting from an instability occurring in protoplanetary disks by \cite{BL1994}, which can be summarized as follows.
The disk is unstable if the rate of heating increases with small perturbations in temperature as compared to the rate of cooling, as this will cause runaway heating of the disk. 
At a lower temperature, the hydrogen in the disk is neutral, and the gas is optically thin at visible and infrared wavelengths.
When the temperature crosses the ionization temperature of hydrogen, the disk suddenly becomes optically thick due to H$^-$ scattering.
The steeply rising opacity as a function of the temperature satisfies the condition for local thermal instability in the disk.
This is demonstrated by the now well-known thermal equilibrium S curve in the $T$--$\Sigma$ plane.
The classical thermal instability takes place in the disk at temperatures of about 7000~K, or corresponding mass transfer rates,
which can easily be achieved in cataclysmic variables.
However, a protoplanetary disk is much cooler, and it is debatable if the disk can get to these hot temperatures. 
It may be possible that the temperature can approach this value very close to the star, at about 0.1 au. 
However, as a consequence, a \cite{SS1973} $\alpha$ value of about $10^{-3}$, which is an order of magnitude lower than that expected in the presence of MRI, is needed to match the timescale of the observed bursts.
Radiative transfer modeling of the disk of the prototype FU Orionis also implies that this outburst extends to 0.5--1 au \citep{Zhu07}.
These have been some of the drawbacks of the thermal instability as a model for outbursts in protoplanetary disks. 
In this section, we will analyze the outbursting behavior in fully MRI active disk simulations, which is analogous to the classical thermal instability, occurring at much lower temperatures.
These outbursts were completely absent in the layered disk models.

\begin{figure}
\centering
\includegraphics[width=0.5\textwidth]{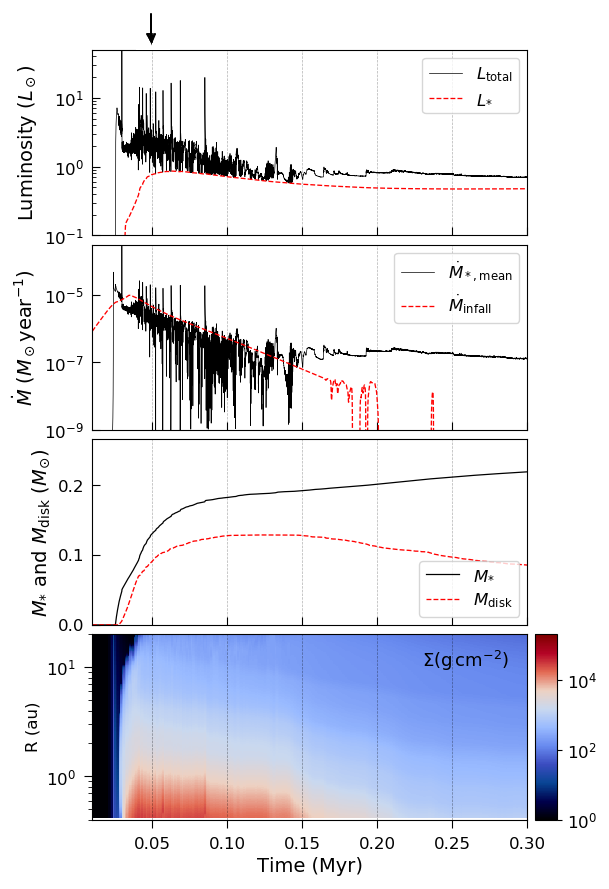}
\caption{Time-dependent quantities -- total and stellar luminosities, mean mass accretion rate, stellar and disk mass -- as well as spacetime diagram of the surface density of the inner 20 au region for the fully MRI-active and low mass model, \simname{model2\_const\_alpha}.
The arrow shows the particular eruption studied in detail.
}
\label{fig:constalltime}
\end{figure}

Figure \ref{fig:constalltime} shows the evolution of global simulation parameters as well as the spacetime diagram of the gas surface density for \simname{model2\_const\_alpha}. 
We chose this low-mass model to demonstrate the thermal instability because the luminosity bursts were not contaminated by clump accretion.
The first panel in Figure \ref{fig:constalltime} shows the luminosity curve where several distinct outbursts can be identified with a peak magnitude between 10 and 20 $L_{\odot}$.
The corresponding mass accretion rate of about $10^{-5} M_\odot {\rm year^{-1}}$ during an outburst can be seen in the second panel.
The eruptive phase of the disk lasted a relatively short time during the early evolution of the disk before about 0.1 Myr, when the surface density in the inner disk was relatively high.
The mass transfer rate stabilized to a few times $10^{-7}$ after this point.
Successive evolution showed a relatively steady and monotonically decreasing accretion rate.
The envelope mass infall rate, measured at 500 au, shows negative values after about 0.15 Myr because of the viscous spread of the disk.
During this early, eruptive phase of the system, the disk was massive and contained over 50\% of the mass of the star at the time.
The spacetime diagram of surface density was monotonically decreasing with radius and showed a much smoother evolution with time. 
The surface density in the inner disk region of a fully MRI-active disk looks very different as compared to that of a magnetically layered disk, as the latter forms a dead zone with dynamical rings.
{ In Table \ref{table:sims}, it is noticeable that the final protostellar mass of \simname{model2\_const\_alpha} was marginally less than that for the corresponding layered disk model (\simname{model2\_T1300\_S100}). 
This can be explained by the fact that a fully magnetically active disk can inhibit the strength of gravitational instability in the initial stages, thus reducing the accretion rate due to gravitational torques \citep{RiceNayakshin18}.}

\begin{figure}
\centering
\includegraphics[width=0.5\textwidth]{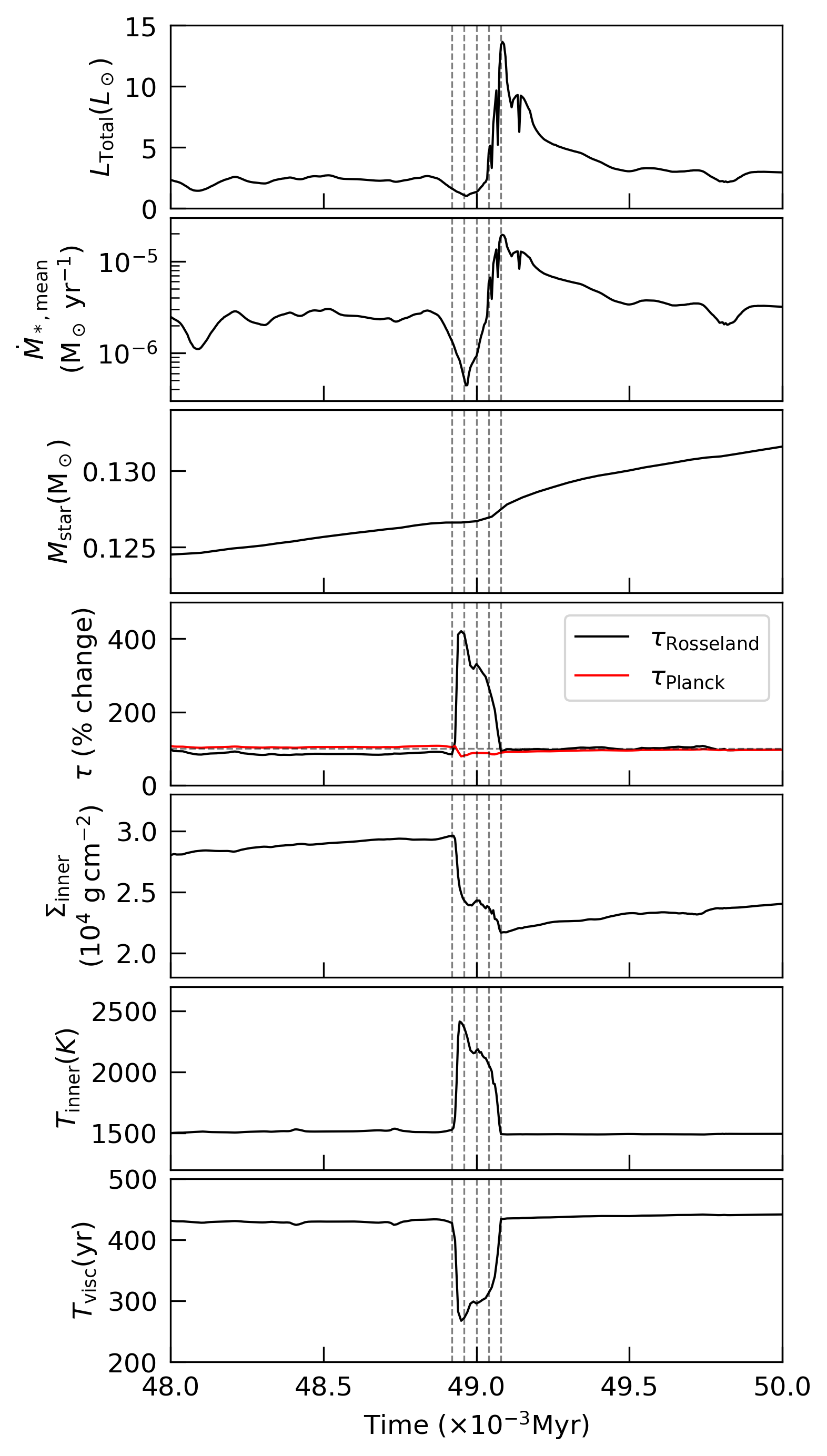}
\caption{Time-dependent system properties (total luminosity, mean mass accretion rate, and stellar mass) as well as Rosseland and Planck opacities, and surface density and midplane temperature at the inner boundary during the thermal instability outburst. The inner boundary was placed at 0.42 au (\simname{model2\_const\_alpha}). The vertical dashed lines mark the time corresponding to the 1D profiles presented in Figure \ref{fig:TI1D}.}
\label{fig:TItimeplot}
\end{figure}

\begin{figure}
\centering
\includegraphics[width=0.5\textwidth]{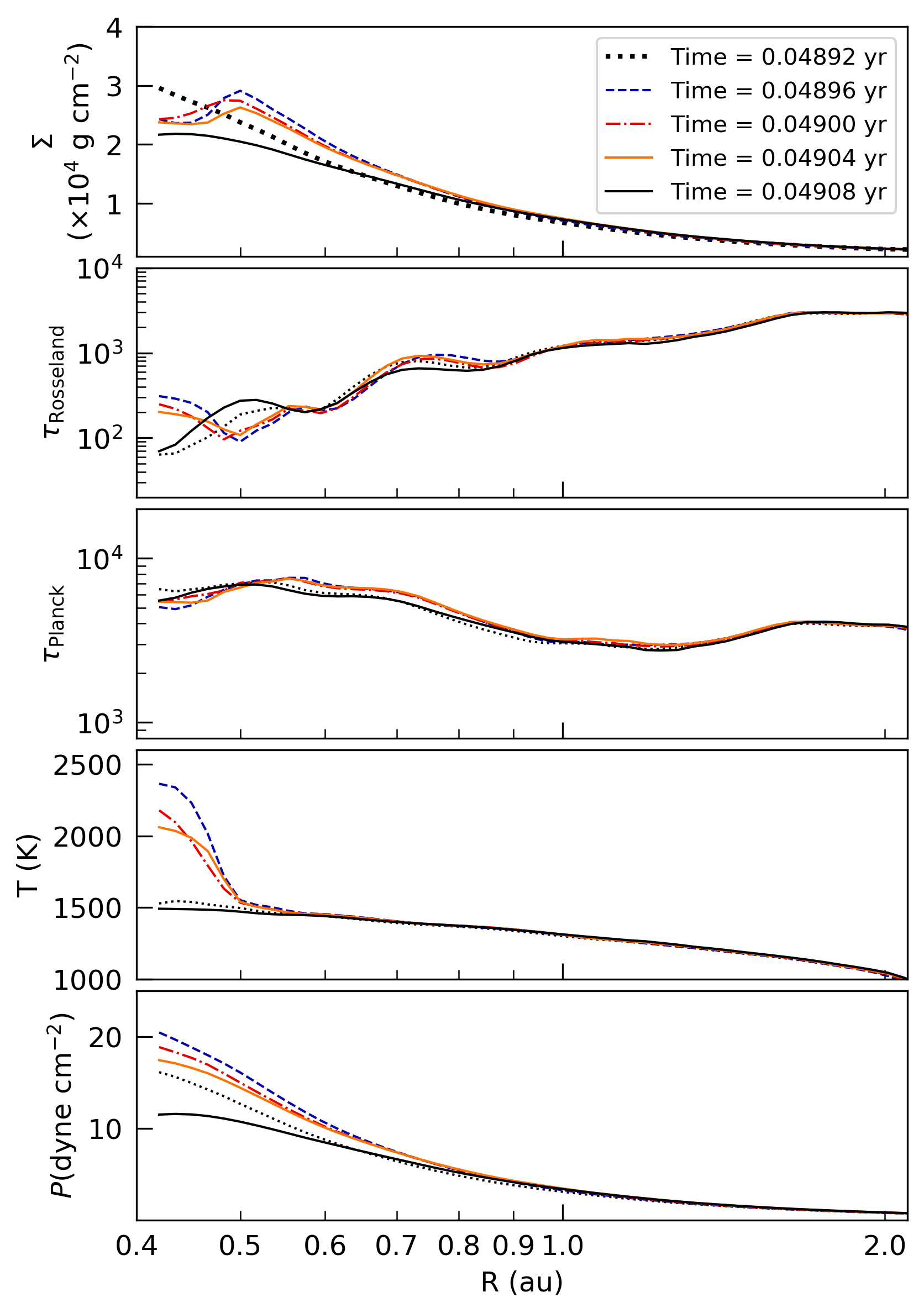}
\caption{Progression of the azimuthally averaged profiles during the thermal outburst, with the inner boundary placed at 0.42 au (\simname{model2\_const\_alpha}).}
\label{fig:TI1D}
\end{figure}

Figure \ref{fig:TItimeplot} focuses on the period of 2000 yr, encompassing a typical mass accretion event occurring at about 0.049 Myr, indicated by an arrow in Figure \ref{fig:constalltime}.
The first panel shows the luminosity outburst with a peak total luminosity of 12 $L_{\odot}$.
The luminosity curve is asymmetric, showing a fast rise over about 50 yr, and a slow decline lasting approximately 300 yr.
The mass transfer rate can be observed to rise from a baseline value of about $2 \times 10^{-6}$ to $2 \times 10^{-5}$ at the maximum.
The estimated amount of accreted material was about 0.0025 $M_\odot$ or about 2 \% of the stellar mass at the time, over the duration of the outburst. 
As seen in the third panel, the stellar mass rose relatively smoothly across the accretion event.
The next three panels in Figure \ref{fig:TItimeplot}, Rosseland and Planck optical depths, surface density, and temperature, all at the inner boundary, give us an insight into the mechanism behind this instability.
The optical depths are calculated as $\tau = \Sigma \kappa/2$, for the respective opacities.
At the inner boundary, they serve as a proxy for the opacity at a given surface density.
The percentage change in the optical depth is calculated with respect to an approximate base value at the beginning of the time period under consideration.
Because the disk at this point is in the optically thick regime ($\tau \gg 1$ in Figure \ref{fig:TI1D}), the Rosseland optical depth dominates the disk cooling.
The Rosseland optical depth increased over four times across the outburst, which implies that the disk cannot cool efficiently over this time period.
The Planck optical depth showed no significant changes.
The midplane temperature at the inner boundary increased from about 1500~K to 2400~K mirroring the Rosseland optical depth.
The slowly increasing surface density at the inner boundary before the outburst reflects the buildup of material in the inner disk region. 
During the outburst, a significant decrease in the surface density was observed as the accumulated material accreted onto the central star.
The increase in temperature coinciding with a similar behavior of Rosseland optical depth across the outburst is consistent with an instability resulting from runaway heating of the disk.
However, because the instability sets in at about 1400~K, it is clear that the outburst was not a result of classical thermal instability which occurs at 7000--10000 K due to ionization of hydrogen.
We call this ``general" thermal instability TI-1.  
{The last row of Figure \ref{fig:TItimeplot} shows the viscous timescale at the inner boundary. 
In this case, viscous timescale reflected only the change in the local sound speed due to the increase in temperature, and its average value across the outburst is consistent with the outburst duration of a few hundred years.  
}

Figure \ref{fig:TI1D} probes the inner disk structure with the azimuthally averaged profiles of quantities -- surface density, Rosseland and Planck optical depths, midplane temperature, and vertically averaged pressure -- in the inner 2 au of the disk during TI-1.
Note that the abscissae are in logarithmic units.
These profiles are aimed at capturing the temporal behavior of the disk at the inner boundary, as opposed to the luminosity burst, and correspond to the vertical dashed lines in Figure \ref{fig:TItimeplot}.
In Figure \ref{fig:TI1D}, the surface density was monotonically decreasing with the radius before the outburst. 
The outburst progressed in an inside-out fashion in the radially outward direction, producing the local maxima seen in surface density profiles. 
The Rosseland optical depth profile changed as seen in the second panel, extending up to 0.5 au, while no significant change in the Planck optical depth was observed. 
When the Rosseland optical depth increased near the inner boundary, the cooling was affected, resulting in the increase in temperature.
The midplane temperature at the inner boundary before the onset of the instability was approximately 1520~K, rising to the maximum value of 2400~K.
A small hysteresis in temperature was after the outburst, as the midplane temperature settled at a value of 1490~K. 
The radial extent of the outburst in terms of temperature was similar to that of the Rosseland optical depth.
The pressure profiles remained monotonically decreasing throughout this time.

\begin{figure}
\centering
\includegraphics[width=0.5\textwidth]{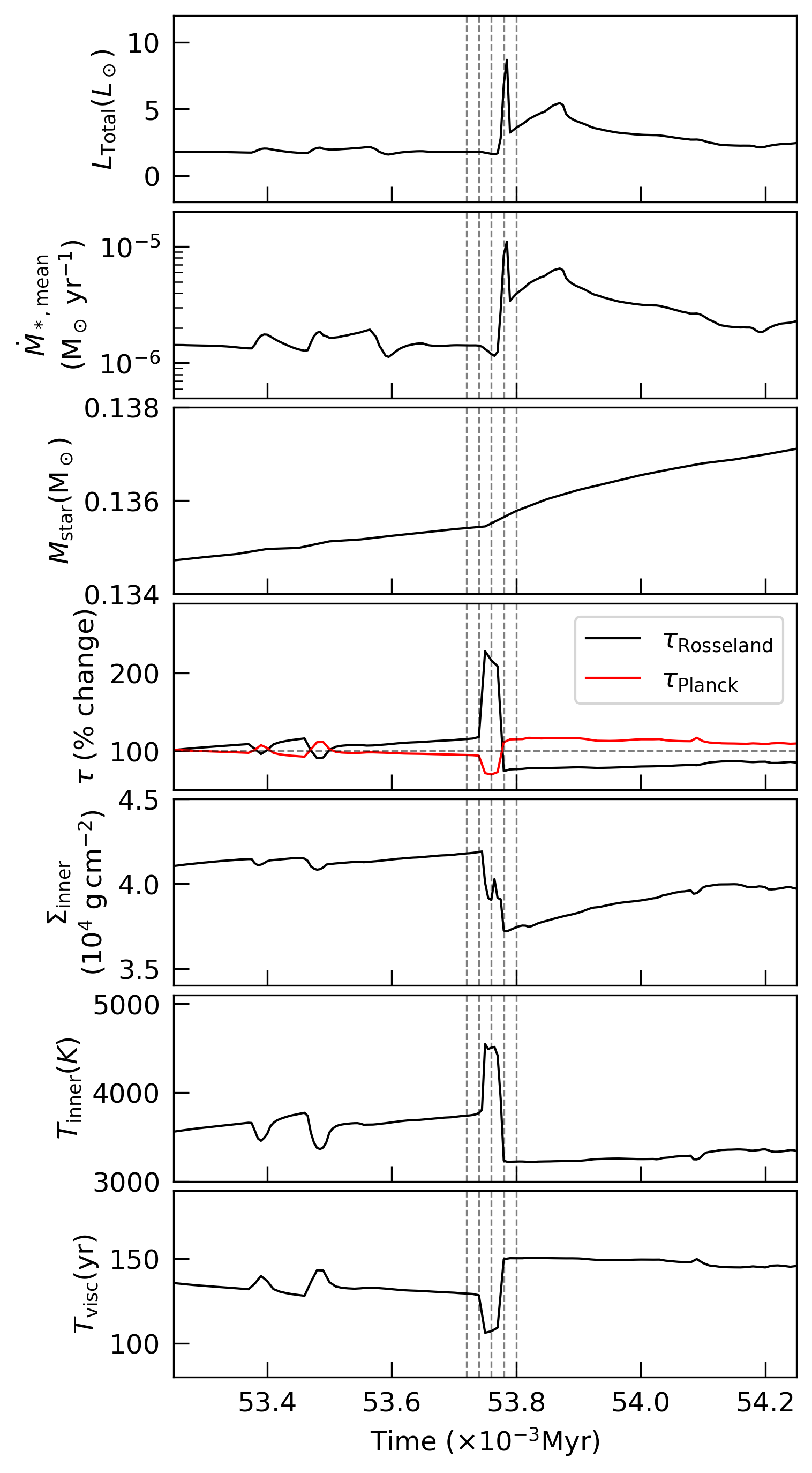}
\caption{Plots for thermal instability, similar to Figure \ref{fig:TItimeplot}, but with the inner boundary placed at 0.21 au (\simname{model2\_const\_alpha\_0.2au}). The vertical dashed lines mark the time corresponding to the 1D profiles presented in Figure \ref{fig:TI1Dsmall}.}
\label{fig:TItimeplotsmall}
\end{figure}

Heretofore the inner boundary of the grid was placed at 0.42 au for all of our investigations. However, the innermost region encompassing the TI-1 outburst (up to 0.5~au) extended about 10 grid cells and thus only marginally resolved. 
In order to study the transition of the disk to a higher temperature state in more detail, we placed the inner boundary of the simulation at 0.21 au, in \simname{model2\_const\_alpha\_0.2au}, keeping all other parameters identical.
Figure \ref{fig:TItimeplotsmall} shows a typical outburst occurring in this simulation, termed TI-2. 
This particular outburst at about 0.0538 Myr was selected because of its proximity to the evolutionary time corresponding to the TI-1 outburst analyzed earlier.
Note that the plot focuses on the 1000 yr surrounding the outburst.
The maximum total luminosity reached about 8~$L_\odot$, with an associated mass accretion rate of $10^{-5}~M_{\odot}$yr$^{-1}$.
{The mean accretion rate and the derived luminosity outburst were bimodal due to the short duration of the burst as well as the rolling window averaging described in Section \ref{sub:analysis}. 
However, the main luminosity outburst lasting about 20 yr was accurately represented.}
The plots of the optical depths, surface density, and temperature at the inner boundary are qualitatively similar to those for the TI-1 outburst. 
Thus, it is clear that the outburst was related to a thermal instability in the inner disk.
However, as compared to TI-1, there were some key differences. 
The peak temperature reached about 4550~K, while the maximum density reached over $4.1\times 10^{-4}$~g~cm$^{-2}$.
The temperature also showed hysteresis with respect to pre- and post-outburst values of about 3750~K to 3200~K.
The peak values of both temperature and density were higher than the TI-1 outburst investigated previously, although still much less than classical thermal instability.
The Rosseland optical depth also showed a relatively small change of about a factor of 2.
{The viscous timescale was shorter in this case because of higher temperature as well as the small radial distance under consideration.} 
These differences suggest that the disk instability, although fundamentally similar, was not identical to TI-1 burst.

\begin{figure}
\centering
\includegraphics[width=0.5\textwidth]{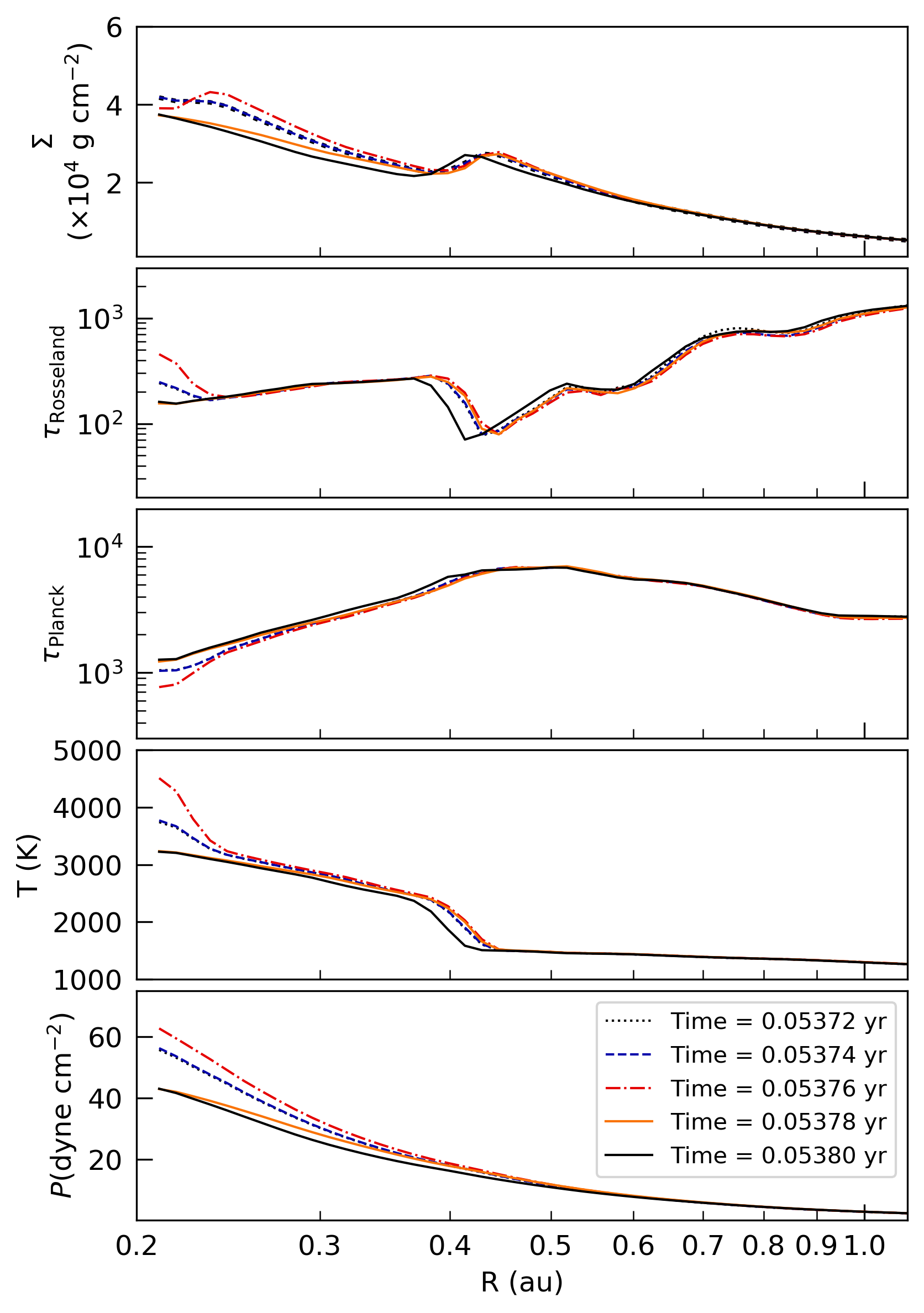}
\caption{Progression of the azimuthally averaged profiles during the thermal outburst, with the inner boundary placed at 0.21 au (\simname{model2\_const\_alpha\_0.2au}).}
\label{fig:TI1Dsmall}
\end{figure}

In Figure \ref{fig:TI1Dsmall} the azimuthally averaged profiles of the quantities -- surface density, Rosseland and Planck optical depths, midplane temperature, and vertically averaged pressure -- are presented for the inner 1 au of the disk during the TI-2 outburst.
The profiles correspond to the instances of time indicated by the vertical dashed lines in Figure \ref{fig:TItimeplotsmall}.
Note that the profiles are sampled at a temporal resolution of 20 yr, as opposed to 40 yr in Figure \ref{fig:TI1D}.
Similar to the TI-1 outburst, the surface density profile showed an inside-out advancement.
The radial extent of the TI-2 outburst was only about 0.25 au. 
The Rosseland optical depth increased across the burst, while the midplane temperature showed a similar trend.
The extent of the outburst was again marginally resolved with about 10 grid cells.

The most notable feature of the radial profiles in Figure \ref{fig:TI1Dsmall} was the structure of the disk at 0.45 au.
Inside of this radial distance, the midplane temperature crossed about 1500~K, and rapidly increased to a higher value of about 2400~K.
These temperature thresholds are similar to the maximum extent in temperature across the previously described TI-1 burst.
A corresponding abrupt increase in Rosseland opacity can also be seen at this radius.
The density showed enhancements similar to those during an outburst, forming a ring at this radius.
The disk structure observed at 0.45 au was secularly stable and did not induce a global instability.
Thus, the disk seems to have continued across the transition in the state associated with TI-1 and found another stable state.
This TI-1 transition region showed inward motion following the decreasing temperature as the disk viscously evolved and cooled down.
There were no associated maxima in pressure, which increased monotonically in the inward direction at all times.
The disk evolved such that the increasing temperature was compensated by the decreasing surface density across the transition region, maintaining a smooth radial pressure profile.

\begin{figure}
\centering
\includegraphics[width=0.5\textwidth]{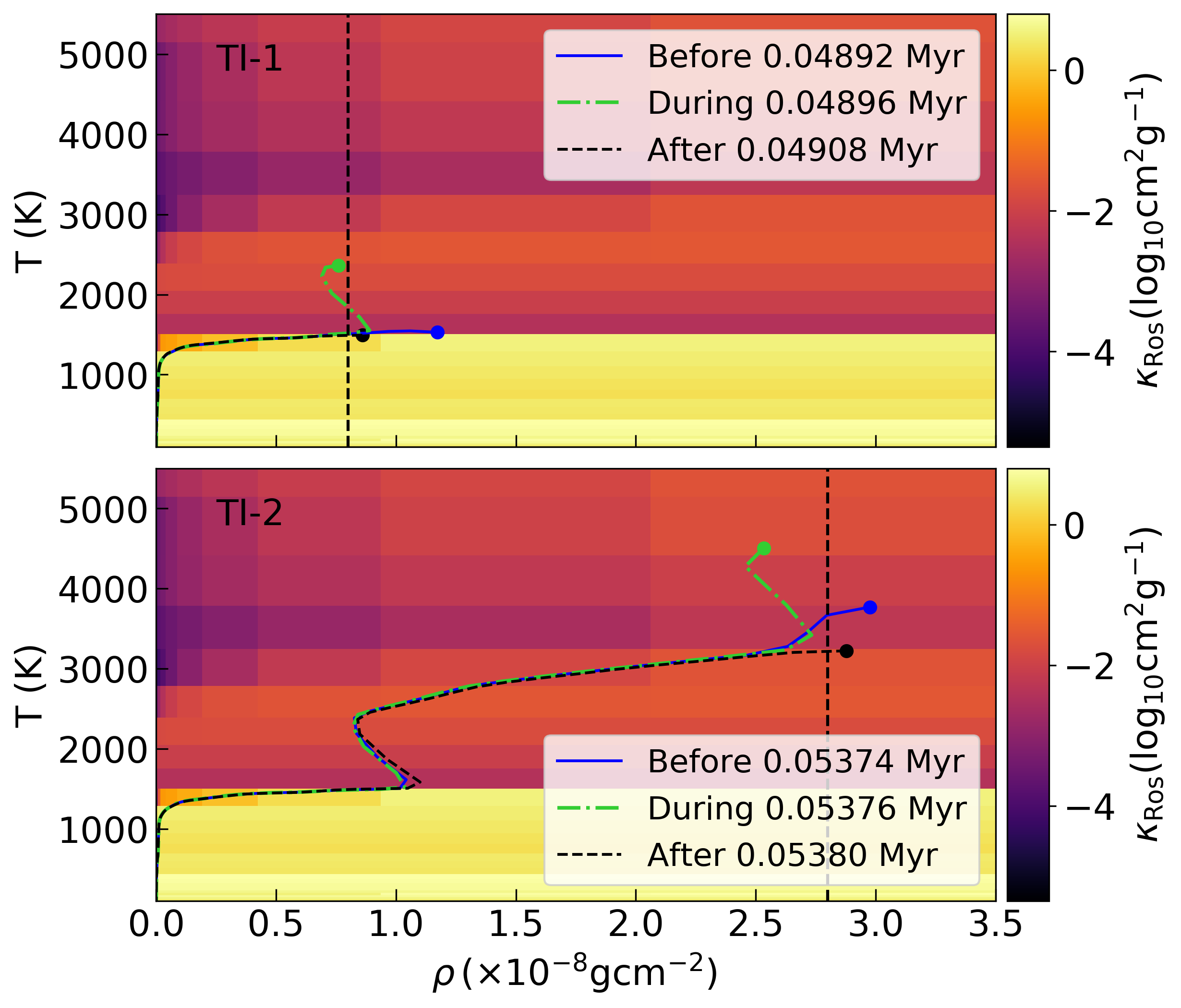}
\caption{Progression of the thermal instabilities in the $\rho$--$T$ plane, with the Rosseland opacity in the background. 
The top and bottom panels show TI-1 and TI-2 with the inner boundary placed at 0.42 (\simname{model2\_const\_alpha}) and 0.21 (\simname{model2\_const\_alpha\_0.2au}) au respectively. The solid points show the value at the inner boundary.
The vertical lines mark density in the vicinity of the instabilities, plotted in Figure \ref{fig:opacitycross}. 
}
\label{fig:rho-T}
\end{figure}

As mentioned earlier, thermal instability arises when the opacity is a rapidly increasing function of temperature.
The Rosseland and Planck mean opacities used in the heating and cooling functions in our model were obtained from the \cite{Semenov2003} opacity tables (see \cite{Kadam2019} for the equations).
In order to investigate the opacity structure, in Figure \ref{fig:rho-T} we plot the azimuthally averaged midplane temperature as a function of central density of the disk, as both the thermal instability outbursts progressed.
The central density was calculated assuming a vertical thermal equilibrium, as this timescale is much shorter than that for the viscous evolution. 
The color diagram in the background shows the Rosseland opacity from the tables in logarithmic units, which is a function of local density and temperature.
The top panel shows progression across TI-1 with the inner boundary of the disk is placed at 0.42 au, while in the bottom panel shows that for TI-2 where the boundary is placed at 0.21 au.
Most of the disk on this $\rho$--$T$ plane lies near the y-axis, at low densities and temperatures.
The filled circles denote the values at the inner boundary, which is canonically the hottest and densest part of a disk. 
The three instances -- before, during and after -- are chosen at shortly before the outburst, at the maximum inner boundary temperature, and after this temperature has returned to the pre-outburst value. 
Thus, Figure \ref{fig:rho-T} shows how the innermost regions of the disk behave through the progression of thermal instability with respect to the local opacity.
During the outburst, the region near the inner boundary of the disk jumps to a state with higher temperature and lower density, as well as a higher value of opacity.
The reverse is true at the end of the outburst, when the disk falls near its original, pre-outburst state.
Noticeable hysteresis was seen in the temperature for TI-2, as it returned to a smaller value at the end.
The most notable feature of TI-2 in the bottom panel was the continuous transition across TI-1 and the secular stability of this region.
The plot suggests that although the disk was locally unstable to TI-1, this did not manifest as a global disk instability.

\begin{figure}
\centering
\includegraphics[width=0.45\textwidth]{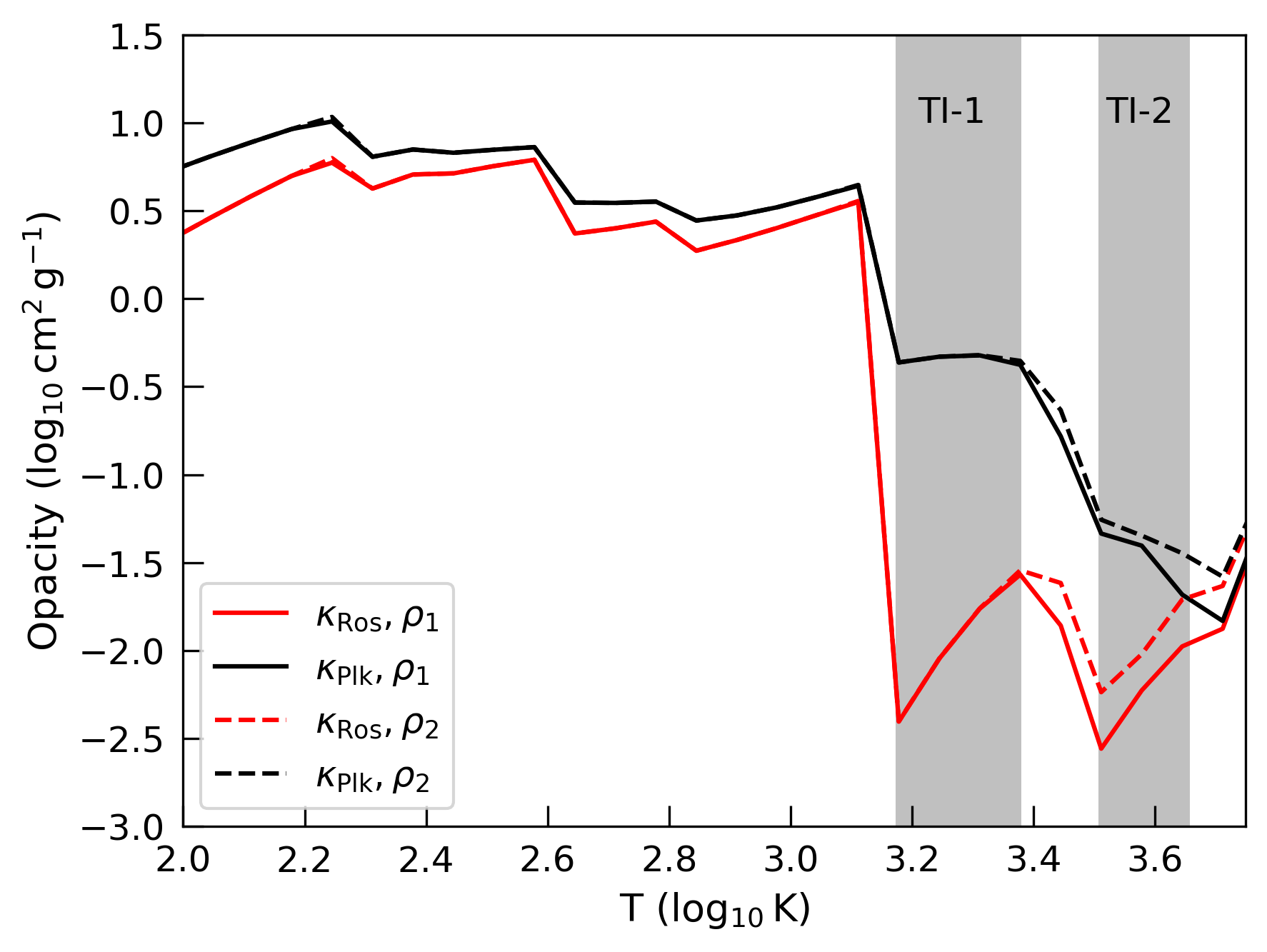}
\caption{The Rosseland and Planck opacities as a function of temperature at a constant density of $\rho_1=0.8$ and $\rho_2=2.8$ $\times 10^{-8}\mathrm{g~cm^{-3}}$.
These densities are chosen in the proximity of the occurrence of the two thermal instability bursts, indicated by vertical dashed lines in Figure \ref{fig:rho-T}.
The increase in midplane temperature across the thermal instabilities is marked by the shaded gray regions, which also coincide with the relatively steep increase in Rosseland opacity.
}
\label{fig:opacitycross}
\end{figure}

In order to investigate the opacity structure in detail, in Figure \ref{fig:opacitycross} we plot the cross section of the opacities at a constant surface density of $\rho_1=0.8$ and $\rho_2=2.8$ $\times 10^{-8} \mathrm{g~cm^{-2}}$ from \cite{Semenov2003}.
These densities are in the vicinity of the two thermal instabilities and are indicated by the vertical dashed lines in Figure \ref{fig:rho-T}.
Both axes are in logarithmic units for direct comparison with previous studies.
The dust opacities dominate below about 1500~K, and above this temperature, dust is sublimated, decreasing the opacity significantly by several orders of magnitude.
Thus, above this temperature, gas opacities come into the picture.
The gray regions in the plot show the two temperatures regimes -- about 1500--2400 K for TI-1 and 3200--4500 for TI-2 -- depicting the extent of temperatures across the thermal instabilities.
Note that the relatively steep dependence of Rosseland opacity on temperature coincides with these regions.
Across the extent of TI-1, the temperature exponent ($\mathrm{d~ln}~\kappa_{\rm Ros}/\mathrm{d~ln}~T_{\rm c}$)
was large and positive at about 4.6, while that across TI-2 was about 4.0.   
{The bump in Rosseland opacity occurring in this region, peaking at about 2500~K is almost entirely due to molecular absorption by { water vapor}, with some contribution from TiO \citep{Alexander94, Semenov2003, Ferguson05}.
The region unstable to TI-1 was terminated near the maximum of this opacity bump.
The region unstable to TI-2 terminated at an inflection point in the opacity curve caused by a combination of molecular lines and continuous opacity from H$^{-}$. 
Opacity due to water was not considered in \cite{BL1994}, and their models showed only classical thermal instability similar to dwarf novae.
Although hydrogen starts to ionize at a few thousand kelvin, the instability sets in at a much higher temperatures of about 5000~K when the temperature exponent becomes large ($\gtrsim7$), and stops when the disk is fully ionized \citep{Lasota2001}.
Note that the exact temperature ranges over which the disk is prone to TI-1 and TI-2 depend on the local values of density.}
The existence of generalized thermal instabilities, TI-1 and TI-2, implies that the chemical composition of the parent cloud core, in particular the water content, can significantly affect the outbursting behavior of a protoplanetary disk via its contribution to the opacity.

\begin{figure}
\centering
\includegraphics[width=0.5\textwidth]{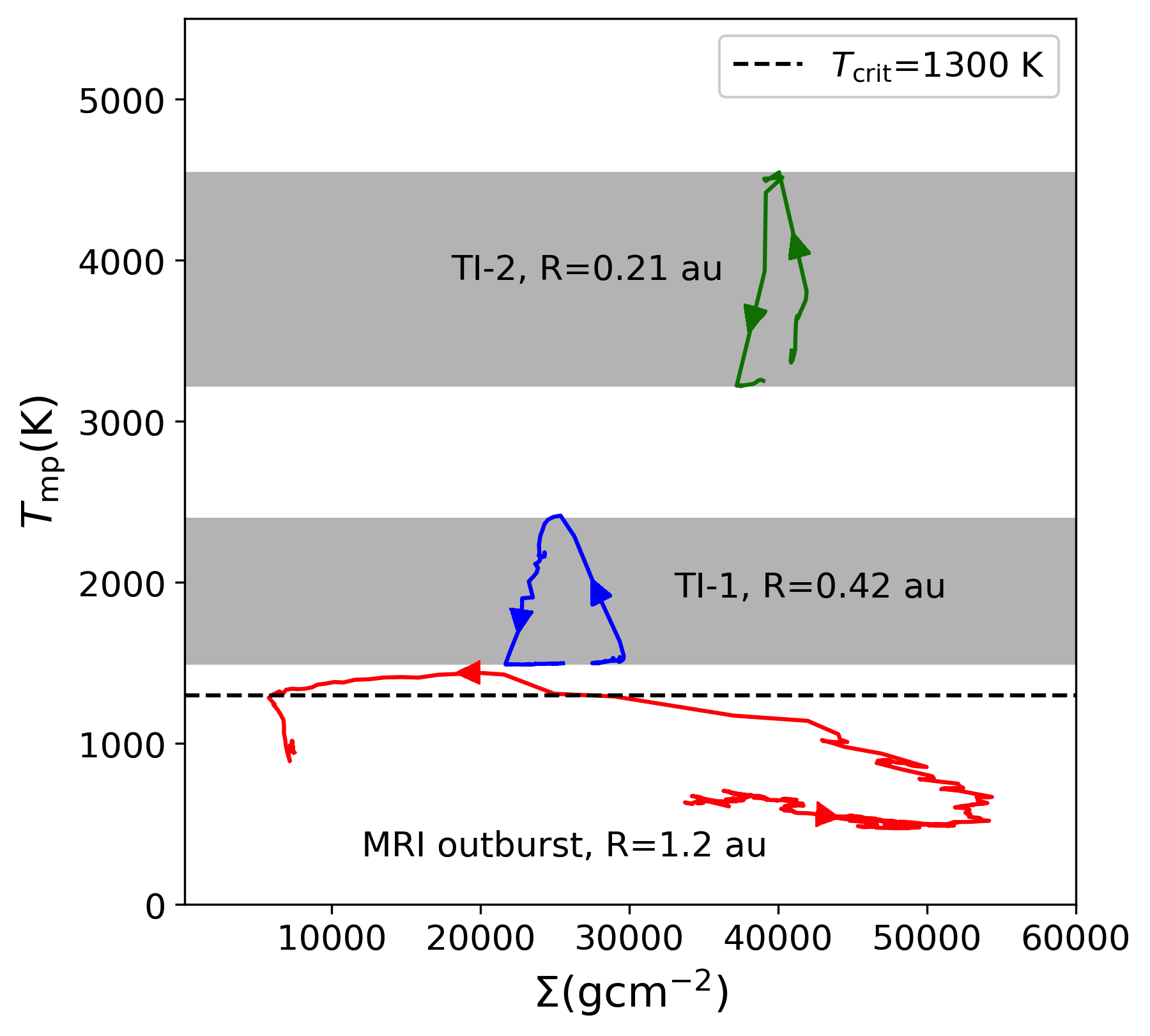}
\caption{The limit-cycle behavior of a disk during MRI outburst as well as TI-1 and TI-2 at a fixed radius of 1.2, 0.42, and 0.21 au, respectively.
The gray area shows the temperature range over which the disk is prone to thermal instabilities, while the dashed line marks the critical temperature for MRI ignition.   
}
\label{fig:limitcycle}
\end{figure}

The classical thermal instability is well understood as an S curve on the $T$--$\Sigma$ plane, where the branches of the stable disk solution are obtained by considering thermal equilibrium. 
There is a region between certain disk temperatures, and consequently the mass transfer rates, that has no steady-state solutions and is unstable.
If the local state of the disk falls within this area in the $T$--$\Sigma$ plane the disk is inherently unstable to thermal perturbations.
This may result in rapid interbranch transitions, and the disk may exhibit time-dependent disk evolution with an associated accretion outburst.
\cite{ML11} demonstrated that the GM instability, where the MRI is triggered by viscous heating in the gravitationally unstable regions, may also be interpreted as an analogous limit cycle.  
Figure \ref{fig:limitcycle} shows the temporal evolution of the state of a disk at a given radius during the two thermal instabilities in this section, as well as the instance of MRI outburst described in Section \ref{sec:mri}.
A radius of 1.2 au was chosen to show the MRI instability, while the values at the inner boundary (0.42 and 0.21 au, respectively) were chosen to best demonstrate TI-1 and TI-2 instabilities.
Consider the curve for MRI instability.
The path traced before the outburst reflects the inward migration of the ring across this radius.
As the location of the ring is cooler than its surroundings due to lower viscous heating in the region, the disk state makes a rightward excursion.
During the MRI outburst occurring above $T_{\rm crit}$, the disk rapidly changes its state to a fully MRI-active branch. 
Here the disk can cool and repeat the limit cycle for MRI outburst.
{During the thermal instabilities shown in Figure \ref{fig:limitcycle}, the disk makes a rapid transition from a lower (cool) to an upper (hot) stable branch, both of which are MRI active.
The regions in between the stable states result from the steep opacity dependence on temperature and thus are unstable to runaway heating.
During an outburst, the mass is drained through accretion on a shorter viscous timescale, and the successive cooling of the disk results in a transition back to the lower branch.  
Note that both TI-1 and TI-2 depended on the location of the inner boundary and due to the limitation of our numerical tools, we could not conclusively establish their global nature.
Further investigations are needed to confirm the outbursting nature of a disk exhibiting such locally unstable annuli.
}

\section{Conclusions and Discussions}
\label{sec:conclusions}

In this study, we investigated the nature of outbursts occurring in the global {hydrodynamics} simulations of magnetically layered disks as well as fully MRI-active disks.
We showed that a typical protoplanetary disk with a layered structure exhibits global disk instability, which results in recurring MRI outbursts.
The outburst duration (a few hundred years), magnitude ($\approx 100 L_\odot$), the peak mass accretion rate ($\approx 10^{-5} M_\odot \, {\rm year}^{-1}$) as well as the radial extent in terms of temperature ($\approx2$ au) are consistent with the FUor eruptions observed in young stellar objects \citep{HK1996, Zhu07, Audard2014}. 
The stellar mass, MRI-triggering temperature, and the thickness of the active layer can all significantly affect the outbursting behavior of a protoplanetary disk.
{ The low-mass model, which ends up with a final protostellar mass of about 0.28 $M_{\odot}$, did not show MRI outbursts, indicating a lower mass limit for observed eruptions similar to FUors.}
At an increased value of ${T_{\rm crit}}$ the MRI outbursts were not triggered, while at a lower value of $\Sigma_{\rm a}$ the eruptions were more powerful.   
Due to the inside-out advancement of the eruption, the rise time of the MRI outburst was much longer than that observed in typical FUors. 
For example, FU Ori and V1057 Cyg have a rise time of the order of a year, while V1515 Cyg shows an order of magnitude slower rising light curve \citep{HK1996}. 
One explanation for this discrepancy may be that the dust physics was not explicitly included in the simulations. 
When the rings are formed, the dust can rapidly accumulate due to pressure maxima in this region.
As the dust dominates the opacity, in its presence the MRI can possibly be triggered in an outside-in fashion, resulting in fast-rising outbursts consistent with the majority of FUor observations.
We are planning to investigate the effects of dust in an upcoming study with the inclusion of a two-part dusty component similar to \cite{Vorobyov2018}.

The fundamental mechanism of the thermal instabilities occurring in fully MRI-active disk simulations was similar to the classical thermal instability due to ionization of hydrogen; 
the rapid increase of opacity with temperature results in thermal disequilibrium and the system exhibits a limit cycle behavior.
The opacity caused primarily by the molecular line absorption by water vapor resulted in a two distinct instabilities -- TI-1 and TI-2 -- starting at about 1500 and 2400~K respectively.
Although both of these thermal instabilities showed associated luminosity bursts in our simulations, it is not clear if a global instability will always ensue when there is such an unstable region within a disk.
In the case of \simname{model2\_const\_alpha}, the inner boundary of the disk was sufficiently close to the region prone to TI-1. 
However, when the boundary was moved closer to the central protostar and away from the unstable region in  \simname{model2\_const\_alpha\_0.2au}, the disk showed a continuous and secularly stable transition across TI-1.
The smooth transition across the unstable region near 0.45 au in Figures \ref{fig:TI1Dsmall} and \ref{fig:rho-T} suggests that the TI-1 instability was only local, while a similar argument can be made for TI-2. 
However, this study offers insights into the nature of these thermal instabilities.
Because TI-1 and TI-2 can occur at much cooler temperatures as compared to classical thermal instability, these are more feasible within the context of protoplanetary disks. 
During low-mass star formation, the accretion disk is typically truncated by the strong magnetic field of the star. 
At such small distances, even when a dead zone is present in the early stages, the disk is fully MRI active due to thermal ionization.
It is possible that in such a scenario, if the locally unstable region encounters this magnetospheric boundary, similar limit cycles will manifest as accretion outbursts via disk-magnetosphere interactions.
The outbursts may be related to TI-2 earlier in the protoplanetary disk's evolution and eventually shift to TI-1 as the disk cools.   
The eruptive phase of the disk will thus depend on its mass (more specifically -- on local temperature and density), water content, and location of the disk truncation radius.  
The viscous timescale 
at the magnetosphere is proportionally shorter and consistent with the observed EXor outbursts.
Thus, TI-1 and TI-2 can possibly serve as a novel mechanism behind some of the shorter duration accretion activity. 
Our current numerical tool cannot probe such small distances, and additional studies are required to determine the behavior of these thermal instabilities conclusively.

We acknowledge the funding from the European Research Council (ERC) under the European Union's Horizon 2020 research and innovation program under grant agreement No. 716155 (SACCRED). E.I.V. acknowledges support from the Austrian Science Fund (FWF) under research grant 
P31635-N27 and Z.R. acknowledges support from the Hungarian OTKA grant No. K-119993.

\acknowledgments

\end{document}